\documentclass[prd,nofootinbib,preprint]{revtex4-1}
\usepackage{amsmath}
\usepackage{amsfonts}
\usepackage{graphicx}
\graphicspath{{Figures/}}
\DeclareGraphicsExtensions{.pdf,.eps}

\numberwithin{equation}{section}

\begin{document}

\title{Non-local tails in radiation in odd dimensions}

\author{M. Khlopunov}
\email{khlopunov.mi14@physics.msu.ru}
\affiliation{Faculty of Physics, Lomonosov Moscow State University, Moscow, 119899, Russia}
\affiliation{Institute of Theoretical and Mathematical Physics, Lomonosov Moscow State University, Moscow, 119991, Russia}

\begin{abstract}
Huygens principle violation in a spacetime of odd dimensions leads to the fact that the retarded massless fields of localised sources depend on their history of motion preceding the retarded time. This non-local character of retarded fields should result into the formation of tail signals in the radiation of localised sources. In particular, in gravity theories with odd number of extra spacetime dimensions the gravitational radiation of binary systems should contain the tail terms. In this work, we demonstrate the presence of tail signal in radiation within a simple model of scalar field interacting with the point charge moving on elliptical orbit in three dimensions. We find that the tail term results into the characteristic dependence of radiation power of the charge on time. In particular, its extremum points do not correspond to the moments when the charge passes the pericenter and apocenter of the orbit, in contrast with the four-dimensional theory. We obtain the formulae for the shifts of radiation power extremum points up to the contributions quadratic in the orbital eccentricity. We also compute the spectral distribution of radiation power of the charge. We find that in three dimensions the charge on elliptical orbit radiates into the lower harmonics of the spectrum, compared to the four-dimensional theory. We conjecture that in higher dimensions the character of spectral distributions is opposite -- the charge mainly radiates into the higher harmonics of the spectrum.
\end{abstract}

\maketitle

\section{Introduction}

Extra spacetime dimensions are an essential ingredient of the number of modern theories of gravity. On one hand, the string theory, being a main model of quantum gravity, predicts the existence of extra dimensions \cite{Green:1987sp}. On the other hand, in the last two decades the number of phenomenological theories of gravity with extra dimensions has been constructed, aimed to solve certain problems of high energy physics \cite{Antoniadis:1990ew,Arkani-Hamed:1998jmv,Antoniadis:1998ig,Randall:1999ee,Randall:1999vf} and cosmology \cite{Dvali:2000hr,Deffayet:2000uy,Deffayet:2001pu,Deffayet:2002sp} (for review see, e.g., \cite{Rubakov:2001kp,Maartens:2010ar,Cheng:2010pt}). Recently, actively developing gravitational-wave astronomy opens up new ways for the experimental study of extra dimensions (for review see, e.g., \cite{Yu:2019jlb,Ezquiaga:2018btd}). Extra dimensions can manifest themselves in gravitational-wave signals in the number of ways: as additional polarisations of gravitational waves \cite{Andriot2017,Khlopunov:2022ubp,Liu:2022cwb}, tower of massive high-frequency Kaluza-Klein modes of gravitational waves \cite{Andriot2017,Barvinsky:2003jf}, additional contributions to the source of gravitational field \cite{Shiromizu:1999wj,Maeda:2003ar,Garcia-Aspeitia:2013jea,Kinoshita:2005nx}, differences between the propagation of gravitational and counterpart electromagnetic signals \cite{Yu:2016tar,Visinelli:2017bny,Lin:2020wnp, Lin:2022hus}, leakage of gravitational waves into extra dimensions \cite{Deffayet:2007kf,Pardo:2018ipy,Corman:2020pyr,Corman:2021avn,Khlopunov:2022jaw}, signal modifications due to the tidal charges of black holes \cite{ Rahman:2022fay,Mukherjee:2022pwd}, as well as the modifications of quasi-normal modes \cite{Chakraborty:2017qve,Mishra:2021waw} and tidal deformabilities of black holes and neutron stars \cite{Chakravarti:2018vlt,Cardoso:2019vof,Chakravarti:2019aup}. Also, it is worth noting here another powerful tool to probe extra dimensions -- the photographs of black hole shadows \cite{Vagnozzi:2019apd,Banerjee:2019nnj,Neves:2020doc,Banerjee:2022jog}.

In particular, the difference between even and odd number of large extra dimensions can manifest in gravitational waves due to the Huygens principle violation in the latter case \cite{hadamard2014lectures,courant2008methods,Ivanenko_book}. The Huygens principle violation consists in the different behaviour of retarded Green's functions of massless fields in even and odd spacetime dimensions. While in even dimensions the signal from an instant flash of the source reaches the observation point in time interval required to propagate to it with the speed of light and decays instantaneously, in odd dimensions the endless tail signal decaying with time is observed. Mathematically, this is due to the fact that in odd dimensions the retarded Green's functions of massless fields are localised not only on the light cone, as in even dimensions, but also inside it. Therefore, in odd spacetime dimensions retarded massless fields propagate with all velocities up to the speed of light. It leads to the fact that at any given spacetime point the retarded field of a localised source depends on the entire history of its motion preceding the corresponding retarded time, in contrast with even dimensions, where it depends only on the source state at the retarded time. This non-local character of retarded fields in odd dimensions should result into the formation of tail signals in the radiation of localised sources. In particular, in theories of gravity with odd number of extra dimensions the gravitational radiation of binary systems of black holes and neutron stars should contain the tail signals.

The presence of tail terms in the radiation of localised sources in odd dimensions has been demonstrated by use of a number of simple models. Within the framework of a scalar field analog of the RS1 model, in Ref. \cite{Barvinsky:2003jf} it has been shown that the signal from the charge at rest living a finite interval of time contains the tail contribution. Analogously, in Ref. \cite{Chu:2021uea} it has been demonstrated that in odd dimensions electromagnetic and gravitational radiation of the source performing harmonic oscillations for a finite interval of time also contain the tail terms. In this paper, our goal is to study the tail signals in gravitational radiation of binary systems in theories with odd number of non-compact extra dimensions. Namely, we demonstrate the presence of tail term in the radiation of point particle moving on elliptical orbit within a simple model of scalar field in three-dimensional Minkowski spacetime. Although this model clearly is not physically viable, it still captures the basic effects associated with the Huygens principle violation, which are expected in the gravitational radiation of elliptical binary systems within realistic models of gravity with odd number of non-compact extra dimensions such as the RS2 and DGP models.

Due to the unusual behaviour of retarded massless fields in odd dimensions, in most of the literature only the problems of radiation in even dimensions were considered \cite{Kosyakov1999,Cardoso:2002pa,Mironov:2006wi,Mironov:2007nk,Cardoso:2007uy,Kosyakov:2008wa} (see, also, \cite{Kosyakov:1992qx,Kosyakov:2007qc,Kosyakov:2018wek}), while the case of odd dimensions was mainly considered in the context of radiation reaction force \cite{Galtsov:2001iv,Kazinski:2002mp,Kazinski:2005gx,Yaremko2007,Shuryak:2011tt,Dai:2013cwa,Harte:2016fru}. As discussed above, in odd dimensions the radiated part of the field contains the non-local tail term analogous to that found by DeWitt and Brehme in curved four-dimensional spacetime \cite{DeWitt:1960fc,Galtsov:2007zz,Barack:2018yvs}. However, while in the latter case the tail term is due to the scattering of gravitational waves on the curvature of spacetime and its computation is rather complicated, in flat odd-dimensional spacetime the tail term is given in the closed analytic form. The tail term can be dealt with by use of the effective field theory approach to the problems of radiation \cite{Porto:2016pyg,Cardoso:2008gn,Birnholtz:2013ffa,Birnholtz:2015hua}. However, being based on the computations in the momentum space, which are irrelevant to the dimensionality of the spacetime, it does not provide us with the information about the structure of radiation field in the wave zone and the role of tail term in the formation of radiation. Recently, it has been shown that the role of tail term in the radiation can be studied in two ways: by Fourier transforming retarded Green's functions over the time coordinate \cite{Chu:2021uea} or by modifying the radiation definition \cite{Galtsov:2020hhn,Galtsov:2021zpb,Khlopunov:2022ubp,Khlopunov:2022jaw}. In this paper, we follow the second approach.

In this work, we use the Rohrlich-Teitelboim approach to radiation \cite{Rohrlich1961,rohrlich2007,Teitelboim1970} (see, also, \cite{Kosyakov:1992qx,Galtsov:2004uqu,Spirin2009,Galtsov:2010tny,Galtsov:2020hhn,Galtsov:2021zpb}), based on the Lorentz-invariant decomposition of the on-shell energy-momentum tensor of retarded field, to extract the emitted part of three-dimensional scalar field. We demonstrate the presence of tail term in the radiation of non-relativistic charge moving along the fixed elliptical orbit. Namely, we find that the dependence of the radiation power of the charge on time has the characteristic form different from the four-dimensional case. In particular, while in four dimensions maxima and minima of the radiation power correspond to the moments of time when the charge passes through the pericenter and apocenter of the orbit correspondingly, in three dimensions they shift in time from this moments forming the tail signal. Also, we compare the spectral distributions of radiation power of the charge on elliptical orbit in three and four spacetime dimensions. Analogously, the spectral distribution in three dimensions has the characteristic feature -- its maximum corresponds to the lower harmonics of the spectrum, compared with the four-dimensional case.

This paper is organised as follows. In Sec. \ref{II}, we discuss the model under consideration, briefly recall the Rohrlich-Teitelboim approach to radiation, obtain the formula for the power of scalar radiation of a non-relativistic charge in three dimensions and provide a simple example of the tail term in the radiation of a charge with Gaussian acceleration. Section \ref{III} is devoted to the computation of radiation power of a non-relativistic charge on elliptical orbit in three dimensions. We calculate it up to the contributions quadratic in the orbital eccentricity and demonstrate the shift of the radiation power extremum points in time and the formation of the tail signal. In Sec. \ref{IV}, we compare the spectral distributions of radiation power of the charge on elliptical orbit in three and four dimensions. We demonstrate the characteristic difference between three and four dimensions consisting in the three-dimensional radiation being emitted into the lower harmonics of the spectrum, compared with the four-dimensional case. In Sec. \ref{V}, we discuss the obtained results.

\section{The Set Up}\label{II}

Our goal is to study the non-local effects in radiation of particle on elliptical orbit in the Minkowski spacetime of odd dimensions. We consider a simple model of the scalar field interacting with the point charge moving along a fixed orbit in three spacetime dimensions. For convenience of comparing the results with the four-dimensional case, we first formulate the problem in the Minkowski spacetime of arbitrary dimension $D=n+1$, and then turn to the cases $D=3$ and $D=4$.

\subsection{Scalar field of a point charge}

Interaction of the massive particle with the massless scalar field in $(n+1)$-dimensional Minkowski spacetime is given by the action
\begin{equation}
\label{eq:s-p_action}
S = - \int d\tau \,  (m + g \varphi(z)) \sqrt{\dot{z}^\alpha \dot{z}_\alpha} + \frac{1}{2\Omega_{n-1}} \int d^{n+1}x \, \partial^\mu \varphi \partial_\mu \varphi, \quad \dot{z}^\alpha = \frac{dz^\alpha}{d\tau},
\end{equation}
where $m$ and $g$ are mass and scalar charge of the particle correspondingly, $z^{\mu}(\tau)$ is the particle's world line parameterised by its proper time, and $\Omega_{n-1}$ is the area of $(n-1)$-dimensional unit sphere
\begin{equation}
\label{eq:sphere_area}
\Omega_{n-1} = \frac{2\pi^{n/2}}{\Gamma(n/2)}.
\end{equation}
We use the Minkowski metric in form $\eta_{\mu\nu}={\rm diag}(1,-1,\ldots ,-1)$.

The action \eqref{eq:s-p_action} yields the scalar field equation of motion
\begin{align}
\label{eq:sc_eq_motion}
& \square \varphi(x) = - \Omega_{n-1} j(x), \\
\label{eq:sc_current}
& j(x) = g \int d\tau \, \delta^{(n+1)}(x-z),
\end{align}
where $\square = \partial^\mu \partial_\mu$ and we have taken into account that $\dot{z}^\alpha \dot{z}_\alpha = 1$. Also, from the action \eqref{eq:s-p_action} we find the canonical energy-momentum tensor of scalar field. Outside the particle's world line, it is written as
\begin{equation}
\label{eq:EMT_sc_gen}
T_{\mu\nu} = \frac{1}{\Omega_{n-1}} \left \lbrack \partial_{\mu} \varphi \partial_{\nu} \varphi - \frac{1}{2} \eta_{\mu\nu} \partial_{\alpha} \varphi \partial^{\alpha} \varphi \right \rbrack, \quad x^\alpha \neq z^\alpha.
\end{equation}
We assume the particle's world line to be fixed and independent of the loss of energy by the particle for radiation.

The retarded solution to the equation of motion \eqref{eq:sc_eq_motion} is given by the integral
\begin{equation}
\label{eq:sq_eq_gen_sol}
\varphi(x) = - \Omega_{n-1} \int d^{n+1}x' \, G_{n+1}(x-x') j(x'),
\end{equation}
where the retarded Green's function is defined by the equation
\begin{align}
\label{eq:gr_fn_eq}
& \square G_{n+1}(x) = \delta^{(n+1)} (x), \\
\label{eq:ret_cond}
& G_{n+1}(x) = 0, \quad x^0 < 0.
\end{align}
In momentum space, it is given by the following integral
\begin{equation}
\label{eq:ret_gr_fn_fourier}
G_{n+1}(x) = - \int \frac{d^{n+1} p}{(2\pi)^{n+1}} \, \frac{e^{-ipx}}{p^2 + i \varepsilon p^0},
\end{equation}
where $p^2 = p^\alpha p_\alpha$ and $\varepsilon = +0$ defines the shift of integrand's poles in complex $p^0$-plane corresponding to the boundary condition \eqref{eq:ret_cond}. Ivanenko and Sokolov \cite{Ivanenko_book} obtained the recurrent relation between the retarded Green's functions in Minkowski spaces of odd dimensions
\begin{align}
\label{eq:odd_gen_GF}
& G_{2\nu+1}(x) = \frac{(-1)^{\nu-1}}{(2\pi)^{\nu-1}} \left( \frac{1}{r} \frac{d}{dr} \right)^{\nu-1} G_{3}(x), \quad \nu \in \mathbb{N}, \\
\label{eq:2+1_GF}
& G_{3}(x) = \frac{\theta(t)}{2\pi} \frac{\theta(x^2)}{\sqrt{x^2}}, \quad t = x^0, \quad r = |\mathbf{x}|.
\end{align}
From Eqs. \eqref{eq:odd_gen_GF} and \eqref{eq:2+1_GF} one finds that the retarded Green's functions in odd dimensions are localised not only on the light cone, but also inside it. Also, they are given by the combination of terms singular on the light cone. However, in Ref. \cite{Galtsov:2020hhn} it was shown that these singularities mutually cancel out and the retarded fields in odd spacetime dimensions have finite values outside the charges' world lines.

\subsection{Rohrlich-Teitelboim approach to radiation}

We use the Rohrlich-Teitelboim approach to radiation \cite{Rohrlich1961,rohrlich2007,Teitelboim1970} (see, also, \cite{Kosyakov:1992qx,Galtsov:2004uqu,Spirin2009,Galtsov:2010tny,Galtsov:2020hhn}) to extract the emitted part of the scalar field and to compute the radiation power of the charge. This approach is based on the Lorentz-invariant decomposition of the on-shell energy-momentum tensor of retarded field and extraction from it the part satisfying certain conditions allowing to associate this part with the energy-momentum of radiation. For this decomposition the covariant retarded quantities are used in the Rohrlich-Teitelboim approach, which are defined as follows.

Let us consider the particle moving along the world line $z^\mu(\tau)$, parameterised by its proper time, with velocity $v^\mu = dz^\mu/d\tau$ in $D$-dimensional Minkowski spacetime. The observation point coordinates are $x^\mu$. Let us construct a light cone in the past with the top at the observation point and denote the point of its intersection with the world line of the particle as $\hat{z}^\mu \equiv z^\mu(\hat{\tau})$. The corresponding moment of proper time $\hat{\tau}$ is called the retarded proper time and defined by the following equation
\begin{equation}
\label{eq:ret_prop_time_eq}
(x^\mu - \hat{z}^\mu)^2 = 0, \quad x^0 \geq \hat{z}^0.
\end{equation}
In what follows, all the hatted quantities will correspond to this moment of time. Based on this, we define three $D$-vectors: null vector $\hat{X}^\mu = x^\mu - \hat{z}^\mu$ directed from the retarded point of world line to the observation point; unit spacelike vector $\hat{u}^\mu$ orthogonal to the particle's velocity at the retarded proper time; null vector $\hat{c}^\mu = \hat{u}^\mu + \hat{v}^\mu$ aligned with vector $\hat{X}^\mu$. In accordance with the definitions above, these vectors satisfy the following equations
\begin{equation}
\label{eq:ret_cov_quan_prop}
\hat{X}^2 = 0, \quad \hat{u}\hat{v} = 0, \quad \hat{u}^2 = - \hat{v}^2 = -1, \quad \hat{c}^2 = 0,
\end{equation}
where $\hat{u}\hat{v} = \hat{u}^\alpha \hat{v}_\alpha$. As a result, we define the Lorentz-invariant distance $\hat{\rho}$ as the scalar product of two introduced vectors
\begin{equation}
\label{eq:Lor-inv_dist_def}
\hat{\rho} \equiv \hat{v}\hat{X}, \quad \hat{X}^\mu = \hat{\rho} \hat{c}^\mu.
\end{equation}
In a Lorentz frame comoving with the particle at the retarded time, it coincides with the spatial distance from the particle to the observation point. If the particle moves inside a bounded region of space, then the Lorentz-invariant distance $\hat{\rho}$ is equivalent to the spatial distance $r = |\mathbf{x}|$ when the observation point is far form this region
\begin{equation}
\hat{\rho} \to r, \quad r \gg |\mathbf{z}|.
\end{equation}

The Rohrlich-Teitelboim approach to radiation is based on the use of the introduced Lorentz-invariant distance $\hat{\rho}$ to expand tensors and determine the wave zone. Namely, in the Rohrlich-Teitelboim approach radiation is determined by the long-range part of the decomposition of the on-shell energy-momentum tensor of retarded field in the inverse powers of $\hat{\rho}$. In $D$ spacetime dimensions, the on-shell energy-momentum tensor of retarded field is expanded as \cite{Teitelboim1970,Kosyakov:1992qx,Kosyakov1999,Galtsov:2004uqu,Kosyakov:2008wa,Spirin2009,Galtsov:2020hhn}
\begin{align}
&T^{\mu\nu} = T^{\mu\nu}_{\rm Coul} + T^{\mu\nu}_{\rm mix} + T^{\mu\nu}_{\rm rad} \\
&T^{\mu\nu}_{\rm Coul} \sim \frac{A^{\mu\nu}}{\hat{\rho}^{2D-4}}, \quad T^{\mu\nu}_{\rm mix} \sim \frac{B^{\mu\nu}}{\hat{\rho}^{2D-5}} + \ldots + \frac{C^{\mu\nu}}{\hat{\rho}^{D-1}}, \quad T^{\mu\nu}_{\rm rad} \sim \frac{D^{\mu\nu}}{\hat{\rho}^{D-2}}.
\end{align}
Here, the first term $T^{\mu\nu}_{\rm Coul}$ is the energy-momentum tensor of deformed Coulomb part of the retarded field. The second term $T^{\mu\nu}_{\rm mix}$ is the mixed part of the decomposition consisting of more than one term in $D>4$ and absent in $D=3$. Finally, the long-range part of the energy-momentum tensor $T^{\mu\nu}_{\rm rad}$ has the properties allowing to associate it with the energy-momentum of radiation:
\begin{itemize}
\item
it is separately conserved $\partial_\mu T^{\mu\nu}_{\rm rad} = 0$, corresponding to its dynamical independence from the other parts;
\item
it is proportional to the direct product of two null vectors $T^{\mu\nu}_{\rm rad} \sim \hat{c}^\mu \hat{c}^\nu$, corresponding to the propagation of resulting energy-momentum flux with the speed of light $\hat{c}_\mu T_{\rm rad}^{\mu\nu} = 0$;
\item
it decays as $T^{\mu\nu}_{\rm rad} \sim 1/r^{D-2}$ and gives positive definite energy-momentum flux through the distant $(D-2)$-dimensional sphere.
\end{itemize}
Therefore, the radiation power of the field in $D$ dimensions can be computed in a standard way as the energy flux determined by $T^{\mu\nu}_{\rm rad}$ through the distant $(D-2)$-dimensional sphere of radius $r$
\begin{equation}
\label{eq:RT_rad_pow_def}
W_{D} = \int \, T_{\rm rad}^{0i} \, n^{i}\, r^{D-2} \, d\Omega_{D-2}, \quad n^i = x^i/r, \quad i = \overline{1, D-1},
\end{equation}
where $n^i$ is the unit spacelike vector in the direction of observation and $d\Omega_{D-2}$ is the angular element on the $(D-2)$-dimensional sphere. This decomposition of the energy-momentum tensor and the expression for the radiation power are valid both in even and odd dimensions with the only difference that in odd dimensions the emitted part of the energy-momentum tensor depends on the entire history of the particle motion preceding the retarded time $\hat {\tau}$, while in even dimensions it depends only on the particle's state at $\hat{\tau}$.

Note that in theories where the energy-momentum tensor is given by a bilinear form of field derivatives, as it is in our scalar field model \eqref{eq:EMT_sc_gen}, one can define the emitted part of the field derivative (in what follows, for brevity, the emitted part of the field) as its leading $\hat{\rho}$-asymptotics, by analogy with the emitted part of the energy-momentum tensor
\begin{equation}
\label{eq:emit_part_field_def}
T \sim \partial \varphi \partial \varphi, \quad T^{\rm rad} \sim 1/\hat{\rho}^{D-2} \quad \Longrightarrow \quad \lbrack \partial \varphi \rbrack^{\rm rad} \sim 1/\hat{\rho}^{(D-2)/2}.
\end{equation}
Thus, to compute the radiation one can find the emitted part of the field and substitute it into the general expression for the field energy-momentum tensor \eqref{eq:EMT_sc_gen}.

\subsection{Radiation of a non-relativistic charge in $D=3$}

Let us obtain an expression for the emitted part of the field of a non-relativistic charge in three spacetime dimensions and the corresponding formula for the radiation power of the charge.

In three dimensions, the retarded solution \eqref{eq:sq_eq_gen_sol} to the equation of motion \eqref{eq:sc_eq_motion} has the following form
\begin{align}
\label{eq:2+1_gen_sol}
& \varphi(x) = - 2 \pi \int d^{3}x' \, G_{3}(x-x') j(x'), \\
& G_{3} (x) = \frac{\theta(t)}{2\pi} \frac{\theta(x^2)}{\sqrt{x^2}}.
\end{align}
Substituting here expression for the current \eqref{eq:sc_current}, we obtain the retarded field in form of an integral over the particle's proper time
\begin{equation}
\label{eq:2+1_sc_int}
\varphi(x) = - g \int_{-\infty}^{\hat{\tau}} \frac{d\tau}{\sqrt{X^2}}, \quad X^\mu = x^\mu - z^\mu,
\end{equation}
where the retarded time $\hat{\tau}$ is determined by the Eq. \eqref{eq:ret_prop_time_eq}.

In accordance with the Rohrlich-Teitelboim approach, to extract the emitted part of the field we find its derivative
\begin{equation}
\label{eq:2+1_der_start_int}
\partial_\mu \varphi = - \left. \frac{g \hat{c}_\mu}{\sqrt{X^2}} \right \vert_{\hat{\tau}-\varepsilon} + g \int_{-\infty}^{\hat{\tau}-\varepsilon} d\tau \, \frac{X_\mu}{(X^2)^{3/2}}.
\end{equation}
Here we obtained the retarded time derivative by differentiating Eq. \eqref{eq:ret_prop_time_eq} and using the relations \eqref{eq:Lor-inv_dist_def}
\begin{equation}
\label{eq:ret_time_der}
\partial_\mu (x^\alpha - \hat{z}^\alpha)^2 = 0 \quad \Longrightarrow \quad \partial_\mu \hat{\tau} = \hat{c}_\mu.
\end{equation}
As by virtue of Eq. \eqref{eq:ret_cov_quan_prop} the local boundary term in Eq. \eqref{eq:2+1_der_start_int} is singular, we introduced here a regularizing term into the upper integration limit $\hat{\tau} \to \hat{\tau} - \varepsilon$, $\varepsilon \to +0$. Also, the integrand in Eq. \eqref{eq:2+1_der_start_int} is singular at the upper integration limit. However, these singularities mutually cancel out, and the resulting retarded field is finite. We eliminate the singular boundary term by integrating by parts the non-local term. Namely, using the relation
\begin{equation}
\frac{1}{(X^2)^{3/2}} = \frac{1}{vX} \frac{d}{d\tau} \frac{1}{\sqrt{X^2}}
\end{equation}
we rewrite the field derivative \eqref{eq:2+1_der_start_int} as
\begin{equation}
\partial_\mu \varphi = - \left. \frac{g \hat{c}_\mu}{\sqrt{X^2}} \right \vert_{\hat{\tau}-\varepsilon} + \frac{g X_\mu}{vX \sqrt{X^2}} \bigg \vert^{\hat{\tau} - \varepsilon}_{-\infty} + g \int_{-\infty}^{\hat{\tau}-\varepsilon} d\tau \left \lbrack \frac{aX - 1}{(vX)^2 \sqrt{X^2}} X_\mu + \frac{v_\mu}{vX \sqrt{X^2}} \right \rbrack.
\end{equation}
As we assume that the charge moves inside a bounded region of space, here in the second boundary term the lower limit vanishes and the upper limit is eliminated by the first boundary term due to the Eq. \eqref{eq:Lor-inv_dist_def}. As a result, field derivative takes the form
\begin{equation}
\label{eq:2+1_der_res_int}
\partial_\mu \varphi = g \int_{-\infty}^{\hat{\tau}} d\tau \left \lbrack \frac{aX - 1}{(vX)^2 \sqrt{X^2}} X_\mu + \frac{v_\mu}{vX \sqrt{X^2}} \right \rbrack.
\end{equation}
Here, for brevity, we omitted the regularizing term in the upper integration limit, implying it in all further calculations.

We extract the emitted part of the field as the leading $\hat{\rho}$-asymptotics of its derivative \eqref{eq:2+1_der_res_int}. For this, we rewrite the vector $X^\mu$ as
\begin{equation}
\label{eq:Z_def}
X^\mu = \hat{\rho} \hat{c}^\mu + Z^\mu, \quad Z^{\mu} = \hat{z}^\mu - z^\mu.
\end{equation}
By use of the Eq. \eqref{eq:Z_def}, we find the leading asymptotics of two terms in the integrand in Eq. \eqref{eq:2+1_der_res_int} as
\begin{align}
& \frac{aX - 1}{(vX)^2 \sqrt{X^2}} X_\mu \to \frac{(a\hat{c}) \hat{c}_\mu}{(v\hat{c})^2 \sqrt{2\hat{\rho}(Z\hat{c})}} \sim \hat{\rho}^{-1/2}, \\
& \frac{v_\mu}{vX \sqrt{X^2}} \to \frac{v_\mu}{v\hat{c} \sqrt{2 \hat{\rho}^3 (Z\hat{c})}} \sim \hat{\rho}^{-3/2}.
\end{align}
Thus, in accordance with the Rohrlich-Teitelboim approach, we find the emitted part of the scalar field of a point charge in three spacetime dimensions as
\begin{equation}
\label{eq:2+1_emit_part}
\lbrack \partial_\mu \varphi \rbrack^{\rm rad} = \frac{g \hat{c}_\mu}{\sqrt{2\hat{\rho}}} {\cal A}, \quad {\cal A} = \int_{-\infty}^{\hat{\tau}} d\tau \, \frac{a\hat{c}}{(v\hat{c})^2 \sqrt{Z\hat{c}}},
\end{equation}
where we introduced the integral amplitude of the radiation energy-momentum flux $\cal A$. The obtained emitted part of the field has asymptotic $\hat{\rho}^{-1/2}$, in accordance with Eq. \eqref{eq:emit_part_field_def}. Also, the emitted part of the field \eqref{eq:2+1_emit_part} is given by the integral over the history of particle motion preceding the retarded time $\tau \leq \hat{\tau}$, which is a characteristic feature for all odd dimensions \cite{Galtsov:2020hhn,Khlopunov:2022ubp,Khlopunov:2022jaw}. It is these integrals over the history of source motion that should result into the formation of tails in radiation in odd spacetime dimensions. Obviously, in the case of a uniform rectilinear motion of the particle, the emitted part of the field vanishes, as expected for radiation. Finally, we find the emitted part of the energy-momentum tensor of the field by substituting Eq. \eqref{eq:2+1_emit_part} into the general expression for the energy-momentum tensor of the scalar field \eqref{eq:EMT_sc_gen}
\begin{equation}
\label{eq:EMT_2+1_gen}
T_{\mu\nu}^{\rm rad} = \frac{g^2 \hat{c}_{\mu} \hat{c}_{\nu} }{4 \pi \hat{\rho}} {\cal A}^{2},
\end{equation}
The obtained emitted part of the energy-momentum tensor satisfies all the criteria for the energy-momentum tensor of radiation from the Rohrlich-Teitelboim approach.

Let us obtain a non-relativistic approximation for the integral amplitude of the radiation energy-momentum flux \eqref{eq:2+1_emit_part}. We make the following assumptions about the motion of the particle:
\begin{itemize}
\item
it is non-relativistic $|\mathbf{v}| \ll 1, \, \forall \tau$;
\item
it moves inside a bounded region of space $|\mathbf{z}| \leq d, \, \forall \tau$ ($d$ is a characteristic size of this region);
\item
the observation point is far from this region $d \ll r$.
\end{itemize}
Therefore, the covariant retarded quantities are expanded in small parameters $|\mathbf{v}|$ and $|\mathbf{z}|/r$ up to the first order as
\begin{align}
\label{eq:htau_non-rel}
&\hat{\tau} \simeq \bar{t} + \mathbf{n}\bar{\mathbf{z}}, \quad \mathbf{n} = \mathbf{x}/r, \\
\label{eq:hrho_non-rel}
&\hat{\rho} \simeq r \left( 1 - \mathbf{n}\bar{\mathbf{v}} - \mathbf{n}\bar{\mathbf{z}}/r \right), \\
\label{eq:hc_non-rel}
&\hat{c}^\mu \simeq \big \lbrace 1 + \mathbf{n}\bar{\mathbf{v}}, \, \mathbf{n} \left( 1 + \mathbf{n}\bar{\mathbf{v}} + \mathbf{n}\bar{\mathbf{z}}/r \right) - \bar{\mathbf{z}}/r \big \rbrace,
\end{align}
where $\bar{t} = t - r$ is the retarded time computed up to the leading contribution, and all the barred quantities correspond to this moment of time. By use of the Eqs. (\ref{eq:htau_non-rel}--\ref{eq:hc_non-rel}) we rewrite the integrand in Eq. \eqref{eq:2+1_emit_part} as
\begin{equation}
\frac{1}{\sqrt{Z\hat{c}}} \simeq \frac{1}{\sqrt{\bar{t} - t'}} \left \lbrack 1 + \mathbf{n}\bar{\mathbf{v}} - \mathbf{n}\mathbf{s} \left( 1  + \mathbf{n}\bar{\mathbf{v}} + \frac{1}{r} \mathbf{n}\bar{\mathbf{z}} \right) - \frac{1}{r} \mathbf{s}\bar{\mathbf{z}} \right \rbrack^{-1/2},
\end{equation}
where we introduced a spacelike vector $\mathbf{s}$ defined as
\begin{equation}
\label{eq:s_non-rel}
\mathbf{s} = \frac{\bar{\mathbf{z}} - \mathbf{z}(t')}{\bar{t} - t'},
\end{equation}
and replaced the particle's proper time with the coordinate time $\tau=t'$, given their equivalence in the non-relativistic limit. At the retarded time $\bar{t}$ the vector $\mathbf{s}$ has a finite value $\lim_{t' \to \bar{t}} \, \mathbf{s} = \bar{\mathbf{v}}$. Rewriting the particle coordinates in terms of its velocity as
\begin{equation}
\mathbf{z}(t) = \int_{t_{i}}^{t} \mathbf{v}(t') dt' + \mathbf{z}(t_{i}),
\end{equation}
where $t_{i}$ is the initial moment of time in the remote past, and using the first mean value theorem we find that the vector $\mathbf{s}$ is of the same order of magnitude as the particle velocity
\begin{equation}
|\mathbf{s}| \sim |\mathbf{v}|, \, \forall t'.
\end{equation}
Therefore, it should be considered as another small parameter in the non-relativistic expansion. As a result, the integrand in the integral amplitude \eqref{eq:2+1_emit_part} up to the first order in small parameters has the form
\begin{equation}
\label{eq:RCQ_contract_non-rel}
\frac{1}{\sqrt{Z\hat{c}}} = \frac{1}{\sqrt{\bar{t} - t'}} \left( 1 - \frac{1}{2} \mathbf{n} ( \bar{\mathbf{v}} - \mathbf{s} ) \right).
\end{equation}
Analogously, we find the expansions of two scalar products in the integral amplitude \eqref{eq:2+1_emit_part} up to the first order as
\begin{equation}
\label{eq:non-rel_kinem_exp}
a\hat{c} \simeq - \mathbf{n} \mathbf{a}, \quad v\hat{c} \simeq 1 + \mathbf{n} ( \bar{\mathbf{v}} - \mathbf{v}).
\end{equation}
Here, we took into account that the higher derivatives of particle coordinates should also be small for the particle to be non-relativistic the entire history of its motion (for detailed proof see, e.g., \cite{Khlopunov:2022ubp}).

As a result, the integral amplitude \eqref{eq:2+1_emit_part} up to the first order in small parameters takes the form
\begin{equation}
{\cal A} \simeq - \int_{-\infty}^{\bar{t}} dt' \, \frac{\mathbf{n} \mathbf{a}}{\sqrt{\bar{t}-t'}}.
\end{equation}
Substituting it into the Eq. \eqref{eq:EMT_2+1_gen} we find a non-relativistic approximation for the emitted part of the energy-momentum tensor
\begin{equation}
T_{\mu\nu}^{\rm rad} = \frac{g^2 \bar{c}_{\mu} \bar{c}_{\nu} }{4 \pi r} \bigg \lbrack \int_{-\infty}^{\bar{t}} dt' \, \frac{\mathbf{n} \mathbf{a}}{\sqrt{\bar{t}-t'}} \bigg \rbrack^2, \quad \bar{c}^{\mu} = \lbrace 1, \mathbf{n} \rbrace.
\end{equation}
Hence, in accordance with the Eq. \eqref{eq:RT_rad_pow_def}, we find the angular distribution of the scalar field radiation power of a non-relativistic charge in three spacetime dimensions as
\begin{equation}
\label{eq:2+1_nr_rad_pow}
\frac{d W_{3}}{d\Omega_{1}} = \frac{g^2}{4 \pi} \bigg \lbrack \int_{-\infty}^{\bar{t}} dt' \, \frac{\mathbf{n} \mathbf{a}}{\sqrt{\bar{t}-t'}} \bigg \rbrack^2.
\end{equation}
As will be shown below, the Eq. \eqref{eq:2+1_nr_rad_pow} gives the finite value of the radiation power of the charge. As was discussed above, the radiation power of a charge in three dimension depends on the entire history of charge motion preceding the retarded time.

\subsection{Radiation of a non-relativistic charge in $D=4$}

Let us also obtain expressions for the emitted part of the field and the radiation power of a non-relativistic particle in four dimensions. Comparing them with their three-dimensional analogs will let us better understand the features of radiation in odd dimensions.

The retarded field of the point charge in four spacetime dimensions is given by the integral
\begin{align}
& \varphi(x) = - 4\pi \int d^4x' \, G_4(x-x') j(x'), \\
& G_4(x) = \frac{\theta(t)}{2\pi} \delta(x^2).
\end{align}
Substituting here the expression for the current \eqref{eq:sc_current} and using the relation
\begin{equation}
\theta(X^0) \delta(X^2) = \frac{\delta(\tau -\hat{\tau})}{2\hat{\rho}}, \quad X^\mu = x^\mu - z^\mu,
\end{equation}
we find the field of the point charge in four dimensions as
\begin{equation}
\varphi(x) = - \frac{g}{\hat{\rho}}.
\end{equation}
Therefore, in four dimensions it depends only on the state of the source at retarded time. Using Eqs. \eqref{eq:ret_prop_time_eq}, \eqref{eq:Lor-inv_dist_def} and \eqref{eq:ret_time_der} we find the field derivative as
\begin{equation}
\partial_\mu \varphi = \frac{g}{\hat{\rho}^2} ( \hat{\rho} (\hat{a}\hat{c}) \hat{c}_\mu + \hat{v}_\mu - \hat{c}_\mu ).
\end{equation}
Finally, we find the emitted part of the field of a point charge in four dimensions as
\begin{equation}
\lbrack \partial_\mu \varphi \rbrack^{\rm rad} = g \frac{\hat{a}\hat{c}}{\hat{\rho}} \hat{c}_\mu.
\end{equation}

From Eqs. (\ref{eq:htau_non-rel}--\ref{eq:hc_non-rel}) and \eqref{eq:non-rel_kinem_exp} we easily find the non-relativistic approximation for the emitted part of the field
\begin{equation}
\lbrack \partial_\mu \varphi \rbrack^{\rm rad} \simeq - g \frac{\mathbf{n}\bar{\mathbf{a}}}{r} \bar{c}_\mu.
\end{equation}
Hence, by use of the Eqs. \eqref{eq:EMT_sc_gen} and \eqref{eq:RT_rad_pow_def} we obtain the scalar field radiation power of a non-relativistic charge in four spacetime dimensions in the form
\begin{equation}
\label{eq:3+1_nr_rad_pow}
\frac{d W_4}{d \Omega_2} = \frac{g^2}{4\pi} (\mathbf{n}\bar{\mathbf{a}})^2.
\end{equation}
Comparing Eqs. \eqref{eq:2+1_nr_rad_pow} and \eqref{eq:3+1_nr_rad_pow} we find that the expressions for the radiation power of a non-relativistic charge in three and four dimensions have a similar form, except for the integral over the history of charge motion in the first case.

\subsection{Tail signal in radiation of a charge with Gaussian acceleration}

As was shown in Refs. \cite{Galtsov:2020hhn} and \cite{Khlopunov:2022ubp} for the radiation of particle on a circular orbit in odd spacetime dimensions, the tail term does not manifests in radiation when the source moves with constant absolute values of velocity and acceleration. Also, in Ref. \cite{Galtsov:2020hhn} it was shown that in the case of ultra-relativistic source the radiation power dependence on the history of source motion is localised on the small interval of proper time preceding the retarded time, effectively eliminating the tail term. Thus, the tail terms should manifest most distinctly in the radiation of non-relativistic sources with kinematic characteristics changing over time. As a proof of principle, we consider the radiation of a charge with acceleration depending on time as the Gaussian function.

Let us consider the non-relativistic charge moving along a straight line in three spacetime dimensions with acceleration depending on time as the Gaussian function
\begin{equation}
\label{eq:Gauss_acc}
\mathbf{a}(t) = \left \lbrace a_0 \exp \left \lbrack - t^2/2\sigma^2 \right \rbrack , 0 \right \rbrace.
\end{equation}
Here $a_0$ is the characteristic value of acceleration and $\sigma$ is the characteristic time interval during which the acceleration is significantly different from zero. The charge remains non-relativistic if $a_0 \sigma \ll 1$. From Eq. \eqref{eq:2+1_nr_rad_pow} we find the radiation power of this charge as
\begin{equation}
W_3^{\rm g}(\bar{t}) = \frac{g^2 a_0^2}{4} \bigg \lbrack \int_{-\infty}^{\bar{t}} dt' \, \frac{\exp \lbrack - {t'}^2 / 2\sigma^2 \rbrack}{\sqrt{\bar{t}-t'}} \bigg \rbrack^2,
\end{equation}
where we have taken into account that in three spacetime dimensions $\mathbf{n} = \lbrace \cos \phi, \sin \phi \rbrace$. For an analogous charge in four dimensions, using Eq. \eqref{eq:3+1_nr_rad_pow} we find its radiation power in form
\begin{equation}
W_4^{\rm g}(\bar{t}) = \frac{g^2 a_0^2}{3} \exp \left \lbrack - \frac{\bar{t}^2}{\sigma^2} \right \rbrack,
\end{equation}
where we assumed that the charge moves along the $x$ axis and have taken into account that in four spacetime dimensions $\mathbf{n} = \lbrace \cos \phi \sin \theta, \sin \phi \sin \theta, \cos \theta \rbrace$.

\begin{figure}[t]
\center{\includegraphics[width=0.6\linewidth]{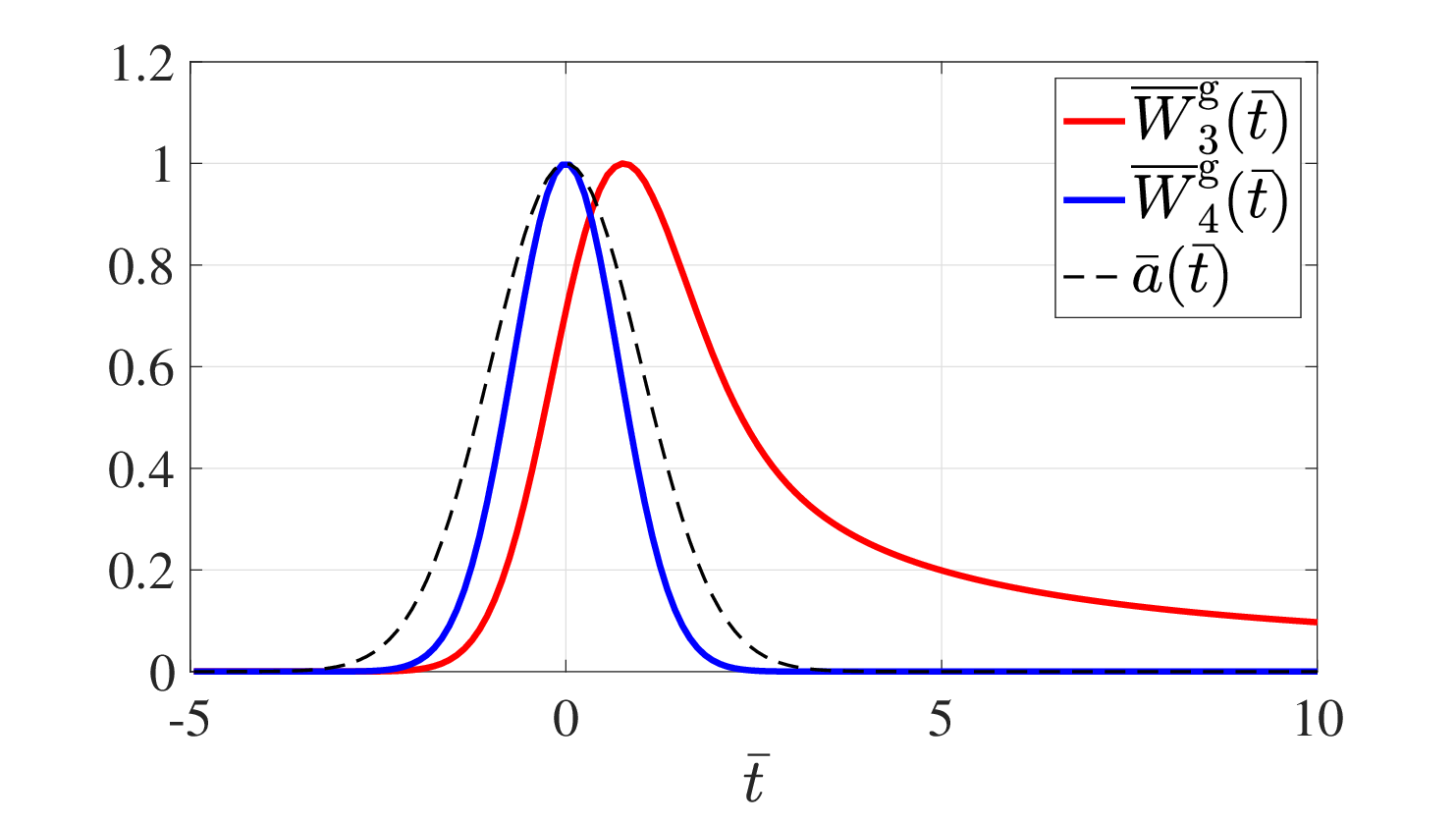}}
\caption{Normalised radiation power of the charge with Gaussian acceleration in three and four spacetime dimensions and the absolute value of its acceleration for $\sigma = 1$. In three dimensions, the long-lived tail signal is present.}
\label{fig:1}
\end{figure}

For convenience of comparing the cases of three and four dimensions, we introduce the radiation power and the absolute value of acceleration normalised to their maximum values
\begin{align}
& \overline{W}_{3}^{\rm g}(\bar{t}) = \frac{W_{3}^{\rm g}(\bar{t})}{W_{\rm 3 max}^{\rm g}}, \quad \overline{W}_{4}^{\rm g}(\bar{t}) = \frac{3}{g^2 a_0^2} W_{4}^{\rm g}(\bar{t}), \quad \overline{a}(\bar{t}) = \exp \left \lbrack - \frac{\bar{t}^2}{2\sigma^2} \right \rbrack, \\
& W_{\rm 3 max}^{\rm g} = \max_{\bar{t}} W_{3}^{\rm g}(\bar{t}).
\end{align}
We numerically obtain their dependence on the retarded time for $\sigma = 1$ (see Fig. \eqref{fig:1}). We find that in three dimensions the long-lived tail signal decaying with time is present, even when the charge acceleration becomes negligible. Also, in three dimensions the maximum of radiation power does not correspond to the moment of time when the charge acceleration takes its maximum value, in contrast to the four-dimensional case.

One can expect an analogous behavior from the radiation of a non-relativistic charge on elliptical orbit in odd spacetime dimensions. However, in this case the radiation of a charge can not acquire such a long-lived tail signal, due to the periodic motion of a charge. Therefore, in the case of a charge on elliptical orbit the most distinct signature of the tail term in radiation should be the shifts of the radiation power extremum points in time from the moments when the charge passes through the pericenter and apocenter of the orbit.

\section{Radiation of a charge on elliptical orbit}\label{III}

Let us consider the radiation of a non-relativistic charge moving along a fixed elliptical orbit in three spacetime dimensions. Dependence of the charge coordinates on time can be written in a parametric form in terms of the eccentric anomaly $\xi$ \cite{Landau_mech}
\begin{align}
\label{eq:2D_ellipt_WL}
& \mathbf{z}(t) = \left \lbrace \rho(t) \cos{\psi(t)}, \rho(t) \sin{\psi(t)} \right \rbrace, \\
\label{eq:rad_pol_ang_on_time}
& \rho(t) = a (1 - e \cos{\xi}), \quad \cos{\psi(t)} = \frac{\cos{\xi} - e}{1 - e \cos{\xi}}, \\
\label{eq:t-anom_rel}
& \omega_0 t = \xi - e \sin{\xi}, \quad \omega_0^2 = \alpha / m a^3,
\end{align}
where $m$ is a mass of the particle, $a$ is a major semiaxis of the ellipse, and $\omega_0$ is a frequency of orbital motion. For simplicity, we assume that the charge motion is driven by some external interaction, which has the Coulomb character with a non-relativistic equation of motion of the charge
\begin{equation}
\label{eq:2D_EoM}
m \mathbf{a} = - \frac{\alpha}{z^3} \mathbf{z}.
\end{equation}
Here $\alpha$ is the constant of this interaction. We do not specify this interaction assuming only that it results into the fixed elliptical orbit of the charge. Also, we neglect the scalar field radiation reaction on the charge motion.

As the eccentric anomaly is a single-valued smooth function of time, it is convenient to pass from the time integral to the integral over anomaly in Eq. \eqref{eq:2+1_nr_rad_pow}. The Jacobian of this transformation is written as
\begin{equation}
\frac{dt}{d\xi} = \frac{1}{\omega_0} (1 - e \cos \xi).
\end{equation}
Thus, we integrate in Eq. \eqref{eq:2+1_nr_rad_pow} over the interval $\xi \in (-\infty,\bar{\xi})$, where the retarded value of anomaly $\bar{\xi}$ is determined from the transcendental equation
\begin{equation}
\label{eq:time_anom_rel}
\omega_0 \bar{t} = \xi - e \sin{\xi} \quad \Longrightarrow \quad \bar{\xi} = \xi(\bar{t}).
\end{equation}
Also, we rewrite $\sin \psi(t)$ in Eq. \eqref{eq:2D_ellipt_WL} as a function of anomaly. By use of the Eq. \eqref{eq:rad_pol_ang_on_time}, we find
\begin{equation}
\label{eq:sin_of_psi_as_anom_func}
\psi(t) = \arccos \frac{\cos{\xi} - e}{1 - e \cos{\xi}} \quad \Longrightarrow \quad \sin{\psi(t)} = \pm \sqrt{1-e^2} \frac{\sin \xi}{1 - e \cos \xi}.
\end{equation}
The charge motion in the direction of increasing polar angle $\psi$ corresponds to the plus sign in Eq. \eqref{eq:sin_of_psi_as_anom_func}. As a result, the charge coordinates as functions of the anomaly take the form
\begin{equation}
\label{eq:WL_of_anom}
\mathbf{z}(\xi) = \left \lbrace a (\cos \xi - e), a \sqrt{1-e^2} \sin \xi \right \rbrace.
\end{equation}
The charge acceleration as a function of the anomaly is found from the equation of motion \eqref{eq:2D_EoM} by substituting the charge coordinates \eqref{eq:WL_of_anom} into it
\begin{equation}
\mathbf{a}(\xi) = - \frac{\omega_0^2 a}{(1 - e \cos \xi)^3} \left \lbrace \cos \xi - e, \sqrt{1-e^2} \sin \xi \right \rbrace.
\end{equation}

As a result, the radiation power of non-relativistic charge on elliptical orbit in three dimensions is given by the following integral
\begin{align}
\label{eq:Rad_pow_gen}
& W_{3} = \frac{g^2 \omega_0^3 a^2}{4\pi} \int d\phi \, J^2 (\bar{\xi},\phi,e), \\
\label{eq:Hist_int_exact}
& J(\bar{\xi},\phi,e) = \int_{-\infty}^{\bar{\xi}} d\xi' \, \frac{\cos \phi (\cos \xi' - e) + \sqrt{1-e^2} \sin \phi \sin \xi'}{\sqrt{\bar{\xi} - \xi' - e(\sin \bar{\xi} - \sin \xi')} (1 - e \cos \xi')^2} \equiv \int_{-\infty}^{\bar{\xi}} d\xi' \, j(\xi',\phi,e),
\end{align}
where we have taken into account that in three spacetime dimensions $\mathbf{n}=\lbrace \cos \phi, \sin \phi \rbrace$. In general, analytical computation of the integral $J(\bar{\xi},\phi,e)$ is complicated. We will calculate it by expanding the integrand $j(\xi',\phi,e)$ in powers of eccentricity.

\subsection{Linear approximation}

Let us calculate the integral over the history of charge motion $J(\bar{\xi},\phi,e)$ in the linear approximation in the orbital eccentricity. This approximation is valid for the eccentricity in range $e \lesssim 10^{-2}$.

The integrand $j(\xi',\phi,e)$ is expanded up to the first order in $e$ as
\begin{align}
& j^{(1)}(\xi',\phi,e) = i^{(0)}(\xi',\phi) + e \, i^{(1)}(\xi',\phi) + {\cal O}(e^2), \\
\label{eq:i^(0)_def}
& i^{(0)}(\xi',\phi) = \frac{\cos \phi \cos \xi' + \sin \phi \sin \xi'}{\sqrt{\bar{\xi} - \xi'}}, \\
\label{eq:i^(1)_def}
& i^{(1)}(\xi',\phi) = \frac{\cos \phi \cos 2\xi' + \sin \phi \sin 2\xi'}{\sqrt{\bar{\xi} - \xi'}} + \frac{1}{2} \frac{(\sin \bar{\xi} - \sin \xi')(\cos \phi \cos \xi' + \sin \phi \sin \xi')}{(\bar{\xi} - \xi')^{3/2}}.
\end{align}
Correspondingly, the integral $J(\bar{\xi},\phi,e)$ up to the first order in eccentricity is written as
\begin{align}
J^{(1)}(\bar{\xi},\phi,e) & = \int_{-\infty}^{\bar{\xi}} d\xi' \, j^{(1)}(\xi',\phi,e) = \int_{-\infty}^{\bar{\xi}} d\xi' \, i^{(0)}(\xi',\phi) + e \int_{-\infty}^{\bar{\xi}} d\xi' \, i^{(1)}(\xi',\phi) + {\cal O}(e^2) \nonumber \\
& = I^{(0)}(\bar{\xi},\phi) + e I^{(1)}(\bar{\xi},\phi) + {\cal O}(e^2).
\end{align}

We start by calculating the integral $I^{(0)}(\bar{\xi},\phi)$
\begin{equation}
I^{(0)}(\bar{\xi},\phi) = \int_{-\infty}^{\bar{\xi}} d\xi' \, i^{(0)}(\xi',\phi) = \int_{-\infty}^{\bar{\xi}} d\xi' \, \frac{\cos \phi \cos \xi' + \sin \phi \sin \xi'}{\sqrt{\bar{\xi} - \xi'}}.
\end{equation}
Transforming the integration variable as $s = \bar{\xi} - \xi'$, we obtain
\begin{equation}
I^{(0)}(\bar{\xi},\phi) = \cos \phi \int_{0}^{\infty} ds \, \frac{\cos(\bar{\xi} - s)}{\sqrt{s}} + \sin \phi \int_{0}^{\infty} ds \, \frac{\sin(\bar{\xi} - s)}{\sqrt{s}}.
\end{equation}
Expanding the cosine/sine of difference and calculating the resulting Fresnel integrals \cite{Zwillinger_table}
\begin{equation}
\label{eq:Fres_int}
\int_{0}^{\infty} ds \, \frac{\cos s}{\sqrt{s}} = \int_{0}^{\infty} ds \, \frac{\sin s}{\sqrt{s}} = \sqrt{\frac{\pi}{2}},
\end{equation}
we find a simple expression for the integral $I^{(0)}(\bar{\xi},\phi)$
\begin{equation}
\label{eq:I^(0)_exp}
I^{(0)}(\bar{\xi},\phi) = \sqrt{\pi} \left \lbrack \sin \bar{\xi} \sin \left( \phi + \frac{\pi}{4} \right) - \cos \bar{\xi} \sin \left( \phi - \frac{\pi}{4} \right) \right \rbrack.
\end{equation}
It coincides with the exact expression for the integral over the history of charge motion \eqref{eq:Hist_int_exact} in the case of charge on a circular orbit $e=0$.

Now we turn to the calculation of $I^{(1)}(\bar{\xi},\phi)$. For convenience, we split the integrand $i^{(1)}(\xi',\phi)$ into two parts with different inverse powers of $\sqrt{\bar{\xi} - \xi'}$
\begin{align}
& i^{(1)}(\xi',\phi) = i^{(1,1)}(\xi',\phi) + i^{(1,3)}(\xi',\phi), \\
& i^{(1,1)}(\xi',\phi) = \frac{\cos \phi \cos 2\xi' + \sin \phi \sin 2\xi'}{\sqrt{\bar{\xi} - \xi'}}, \\
& i^{(1,3)}(\xi',\phi) = \frac{1}{2} \frac{(\sin \bar{\xi} - \sin \xi')(\cos \phi \cos \xi' + \sin \phi \sin \xi')}{(\bar{\xi} - \xi')^{3/2}}.
\end{align}
Accordingly, $I^{(1)}(\bar{\xi},\phi)$ also splits into two integrals
\begin{align}
& I^{(1)}(\bar{\xi},\phi) = I^{(1,1)}(\bar{\xi},\phi) + I^{(1,3)}(\bar{\xi},\phi), \\
& I^{(1,1)}(\bar{\xi},\phi) = \int_{-\infty}^{\bar{\xi}} d\xi' \, i^{(1,1)}(\xi',\phi), \quad I^{(1,3)}(\bar{\xi},\phi) = \int_{-\infty}^{\bar{\xi}} d\xi' \, i^{(1,3)}(\xi',\phi).
\end{align}
The integral $I^{(1,1)}(\bar{\xi},\phi)$ is calculated by analogy with $I^{(0)}(\bar{\xi},\phi)$ and takes the form
\begin{equation}
I^{(1,1)}(\bar{\xi},\phi) = \sqrt{\frac{\pi}{2}} \left \lbrack \sin 2\bar{\xi} \sin \left( \phi + \frac{\pi}{4} \right) - \cos 2\bar{\xi} \sin \left( \phi - \frac{\pi}{4} \right) \right \rbrack.
\end{equation}
Computing $I^{(1,3)}(\bar{\xi},\phi)$, we first integrate it by parts to rewrite it as a combination of Fresnel integrals
\begin{align}
I^{(1,3)}(\bar{\xi},\phi) & = \int_{-\infty}^{\bar{\xi}} d\xi' \, \frac{1}{2} \frac{(\sin \bar{\xi} - \sin \xi')(\cos \phi \cos \xi' + \sin \phi \sin \xi')}{(\bar{\xi} - \xi')^{3/2}} \nonumber \\
& = \sqrt{\varepsilon} \cos \bar{\xi} ( \cos \phi \cos \bar{\xi} + \sin \phi \sin \bar{\xi} ) + \int_{-\infty}^{\bar{\xi}} \frac{d\xi'}{\sqrt{\bar{\xi} - \xi'}} \big \lbrack \cos \xi' ( \cos \phi \cos \xi' \nonumber \\
& + \sin \phi \sin \xi' ) - ( \sin \bar{\xi} - \sin \xi' ) \left( \sin \phi \cos \xi' - \cos \phi \sin \xi' \right) \big \rbrack.
\end{align}
Here, in the first line, we implicitly introduced the regularising term into the upper integration limit $\bar{\xi} \to \bar{\xi} - \varepsilon$, $\varepsilon \to + 0$. The boundary term vanishes and after some transformations $I^{(1,3)}(\bar{\xi},\phi)$ is written as
\begin{equation}
I^{(1,3)}(\bar{\xi},\phi) = I^{(1,1)}(\bar{\xi},\phi) + \cos \phi \sin \bar{\xi} \int_{-\infty}^{\bar{\xi}} d\xi' \, \frac{\sin \xi'}{\sqrt{\bar{\xi} - \xi'}} - \sin \phi \sin \bar{\xi} \int_{-\infty}^{\bar{\xi}} d\xi' \,\frac{\cos \xi'}{\sqrt{\bar{\xi} - \xi'}}.
\end{equation}
The remaining integrals are reduced to the Fresnel integrals \eqref{eq:Fres_int} and $I^{(1,3)}(\bar{\xi},\phi)$ takes the form
\begin{equation}
I^{(1,3)}(\bar{\xi},\phi) = I^{(1,1)}(\bar{\xi},\phi) - \sqrt{\pi} \left \lbrack \frac{1}{2} \sin 2\bar{\xi} \sin \left( \phi + \frac{\pi}{4} \right) + \sin^2 \bar{\xi} \sin \left( \phi - \frac{\pi}{4} \right) \right \rbrack.
\end{equation}
Thus, combining $I^{(1,1)}(\bar{\xi},\phi)$ and $I^{(1,3)}(\bar{\xi},\phi)$ we find the integral $I^{(1)}(\bar{\xi},\phi)$ as
\begin{align}
\label{eq:I^(1)_exp}
I^{(1)}(\bar{\xi},\phi) & = \left( \sqrt{2\pi} - \frac{1}{2} \sqrt{\pi} \right) \sin 2\bar{\xi} \sin \left( \phi + \frac{\pi}{4} \right) - \left( \sqrt{2\pi} - \frac{1}{2} \sqrt{\pi} \right) \cos 2\bar{\xi} \sin \left( \phi - \frac{\pi}{4} \right) \nonumber \\ & - \frac{1}{2} \sqrt{\pi} \sin \left( \phi - \frac{\pi}{4} \right).
\end{align}

\begin{figure}[t]
\center{\includegraphics[width=0.6\linewidth]{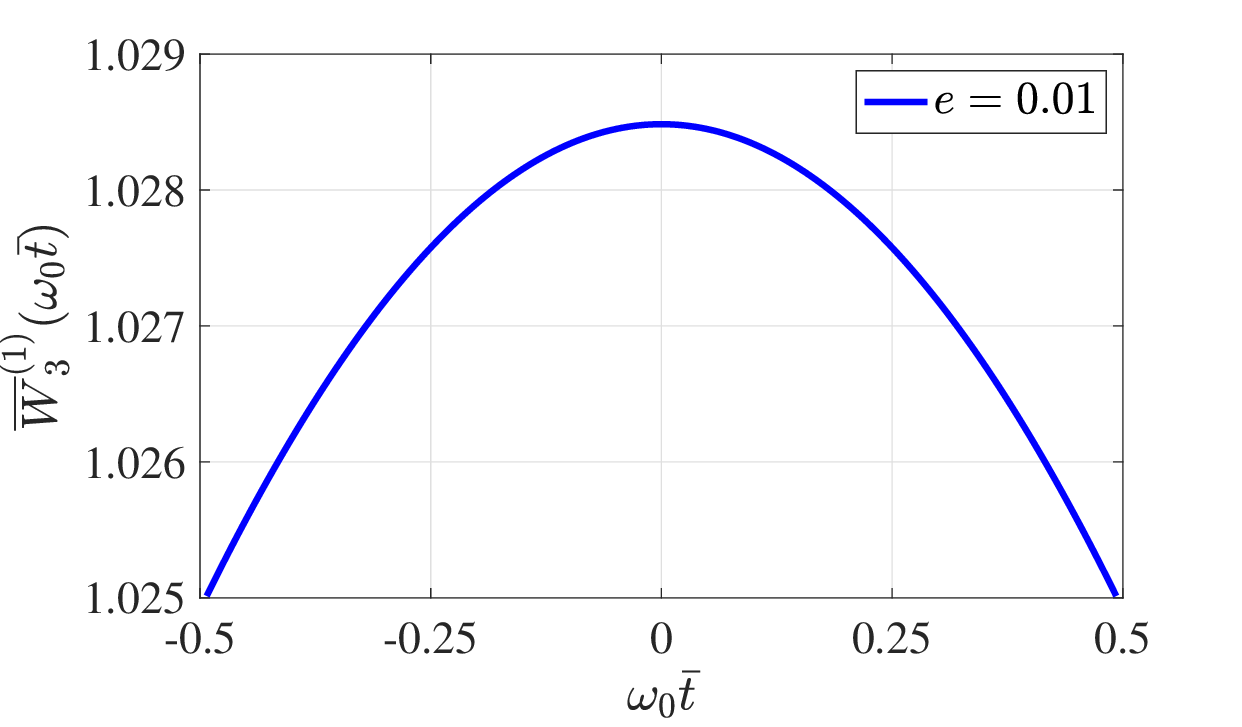}}
\caption{Normalised radiation power of the charge on elliptical orbit in three spacetime dimensions in the linear approximation in orbital eccentricity for $e=0.01$. The maximum of radiation power corresponds to the moment when the charge passes through the pericenter of the orbit $\bar{t}=0$.}
\label{fig:2}
\end{figure}

Finally, the integral over the history of charge motion $J(\bar{\xi},\phi,e)$ up to the first order in orbital eccentricity is given by
\begin{align}
& J^{(1)}(\bar{\xi},\phi,e) = A(\bar{\xi},e) \sin \left( \phi + \frac{\pi}{4} \right) - B(\bar{\xi},e) \sin \left( \phi - \frac{\pi}{4} \right), \\
& A(\bar{\xi},e) = \sqrt{\pi} \left \lbrack \sin \bar{\xi} + e \Big( \sqrt{2} - \frac{1}{2} \Big) \sin 2\bar{\xi} \right \rbrack, \\
& B(\bar{\xi},e) = \sqrt{\pi} \left \lbrack \cos \bar{\xi} + e \Big( \sqrt{2} - \frac{1}{2} \Big) \cos 2\bar{\xi} + e \frac{1}{2} \right \rbrack.
\end{align}
Calculating the angular integrals in Eq. \eqref{eq:Rad_pow_gen}, we find the radiation power of the non-relativistic charge on elliptical orbit in three spacetime dimensions as
\begin{equation}
\label{eq:Rad_pow_1st}
W_{3}^{(1)}(\bar{\xi}) = \frac{\pi}{4} g^2 \omega_0^3 a^2 \left \lbrack 1 + 2 \sqrt{2} e \cos \bar{\xi} + ( \sqrt{2} - 1/2 ) e^2 \cos 2\bar{\xi} + (5/2 - \sqrt{2}) e^2 \right \rbrack.
\end{equation}
Although it contains the terms quadratic in orbital eccentricity, we call it, for brevity, the linear approximation of radiation power, as it is determined by the linear approximation of the integral over the history of charge motion. Normalizing the radiation power to its value for a circular orbit
\begin{equation}
\label{eq:Rad_pow_1st_norm}
\overline{W}_{3}^{(1)}(\bar{\xi}) = \frac{4}{\pi g^2 \omega_0^3 a^2} W_{3}^{(1)}(\bar{\xi}),
\end{equation}
by use of the Eq. \eqref{eq:time_anom_rel} we obtain its dependence on the retarded time (see Fig. \eqref{fig:2}). From Fig. \eqref{fig:2} one finds that in the linear approximation the tail contribution in radiation is negligible. This result could be expected, given that the linear approximation is valid for the orbits extremely close to a circular orbit $e \lesssim 10^{-2}$.

To verify this, we find the radiation power extremum points
\begin{equation}
 \frac{d\overline{W}_{3}^{(1)}}{d\bar{\xi}} = - 2 e \sin \bar{\xi} \left \lbrack \sqrt{2} + ( 2\sqrt{2} - 1 ) e \cos \bar{\xi} \right \rbrack = 0.
\end{equation}
Part of them correspond to the moments when the charge passes through the pericenter and apocenter of the orbit
\begin{equation}
\bar{\xi} = \pi n, \; n \in \mathbb{Z} \quad \Longrightarrow \quad \omega_0 \bar{t} = \pi n, \; n \in \mathbb{Z}.
\end{equation}
The remaining extremum points are determined from the equation
\begin{equation}
\sqrt{2} + ( 2\sqrt{2} - 1 ) e \cos \bar{\xi} = 0,
\end{equation}
which has a solution only if the following inequality is satisfied
\begin{equation}
\label{eq:ecc_crit_1st}
\frac{\sqrt{2}}{( 2\sqrt{2} - 1 ) e} \leq 1 \quad \Longleftrightarrow \quad e \geq e_{\rm cr} = \frac{\sqrt{2}}{( 2\sqrt{2} - 1 )} \simeq 0.77.
\end{equation}
Thus, there are additional extremum points only if the orbital eccentricity exceeds the critical value $e_{\rm cr}$. Obviously, in the range of applicability of linear approximation the inequality \eqref{eq:ecc_crit_1st} does not hold. Therefore, since the radiation power \eqref{eq:Rad_pow_1st} is an even function of anomaly, and since the extremum points of the radiation power correspond to the moments when the charge passes through the pericenter and apocenter of the orbit, the tail term does not manifest itself in radiation in the linear approximation in the orbital eccentricity.

\subsection{Numerical calculations -- linear approximation is not accurate}

From the linear approximation of the radiation power of a charge on elliptical orbit \eqref{eq:Rad_pow_1st} we obtained the critical eccentricity $e_{\rm cr} \simeq 0.77$, above which the tail term starts to manifest in the radiation. To approximate the integral $J(\bar{\xi},\phi,e)$ by the series in powers of eccentricity in the region $e \sim e_{\rm cr}$, higher orders of series expansion are required, which make analytical computations complicated. However, the linear approximation prediction for the critical eccentricity is not accurate and the tail term manifests in the radiation at much lower values of the orbital eccentricity.

\begin{figure}[t]
\begin{minipage}[h]{0.49\linewidth}
\center{\includegraphics[width=1\linewidth]{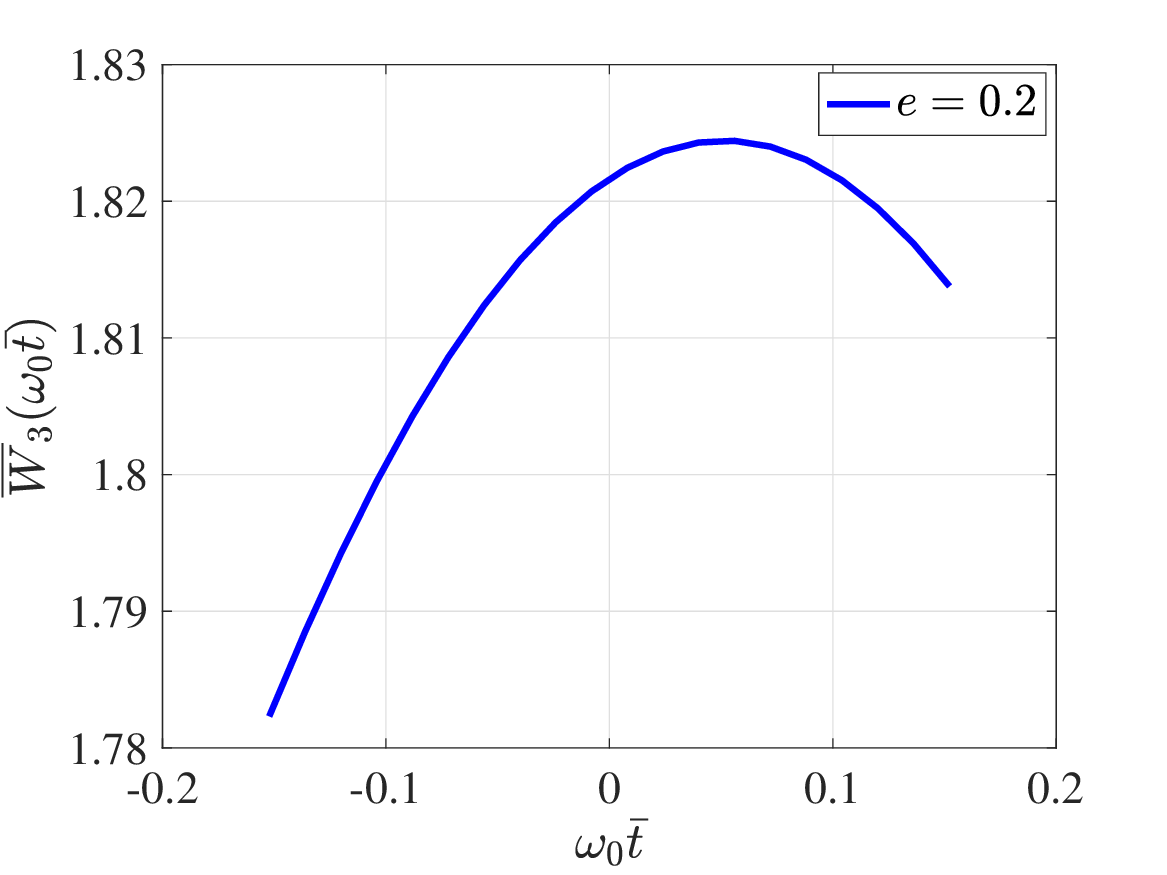}}
\end{minipage}
\hfill
\begin{minipage}[h]{0.49\linewidth}
\center{\includegraphics[width=1\linewidth]{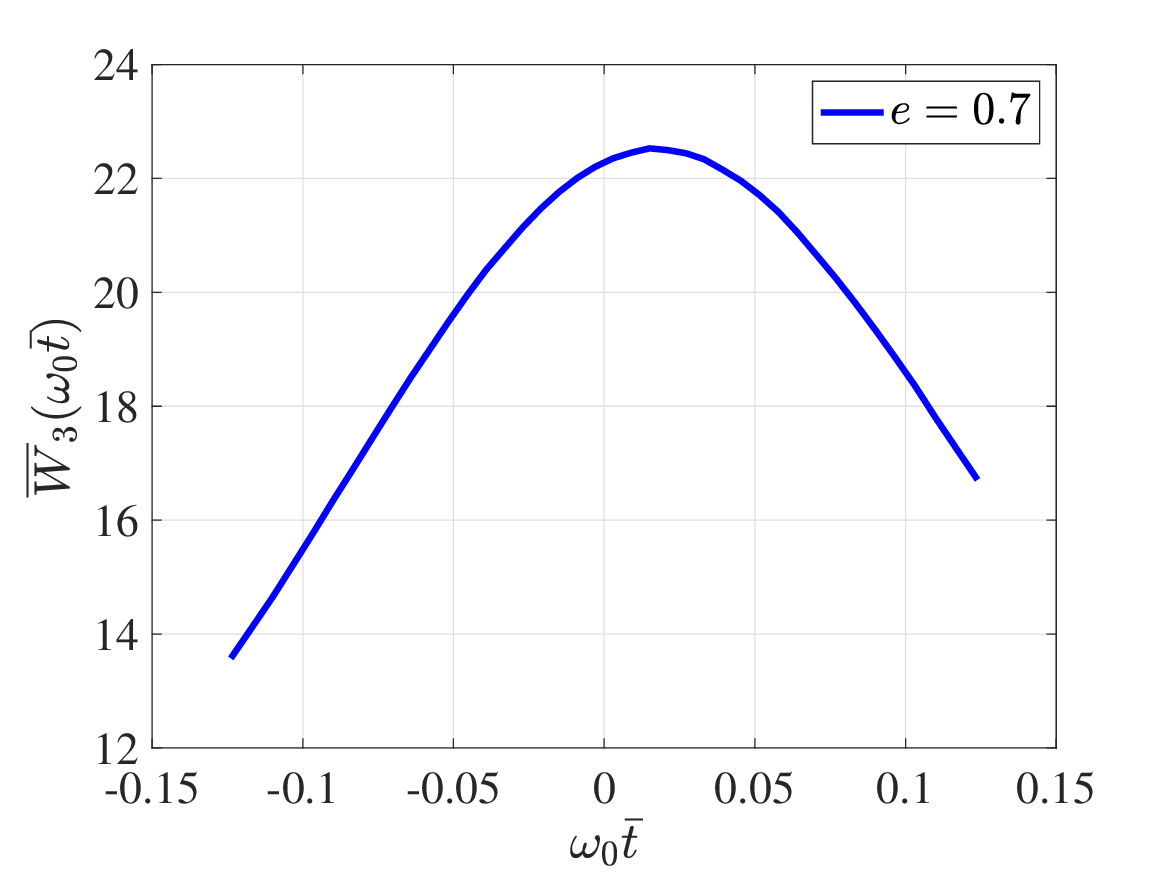}}
\end{minipage}
\caption{Numerical plots of normalised radiation power of the charge on elliptical orbit in three spacetime dimensions for $e=0.2, \, 0.7$. The maximums of radiation power are shifted in time from the moments when the charge passes through the pericenter of the orbit $\bar{t}=0$, forming the tail signals.}
\label{fig:3}
\end{figure}

To demonstrate this, we numerically obtain the dependence of radiation power of the charge on the retarded time for various values of the orbital eccentricity. By analogy with the Eq. \eqref{eq:Rad_pow_1st_norm}, we introduce the normalized radiation power of the charge
\begin{equation}
\overline{W}_{3}(\bar{\xi}) = \frac{4}{\pi g^2 \omega_0^3 a^2} W_{3}(\bar{\xi}),
\end{equation}
where $W_{3}(\bar{\xi})$ is determined from the Eqs. \eqref{eq:Rad_pow_gen} and \eqref{eq:Hist_int_exact}. It is found that the tail term distinctly manifests in radiation already for $e \sim 0.1$ (see Fig. \eqref{fig:3}). In this range of eccentricity values, the integral over the history of charge motion $J(\bar{\xi},\phi,e)$ can be approximated by the series in eccentricity up to the second order. Thus, in this range of eccentricity values the tail term in radiation can be studied analytically.

By analogy with the tail contribution into the radiation of a charge with Gaussian acceleration, the characteristic signature of the tail term in radiation of a charge on elliptical orbit is shifts of the radiation power extremum points from the moments when charge acceleration takes its minimum/maximum values -- moments when charge passes through the apocenter and pericenter of the orbit correspondingly.

\subsection{Quadratic approximation}

Let us calculate the radiation power of a charge on elliptical orbit up to the second order in eccentricity. This approximation is valid for the orbits with eccentricity $e \lesssim 0.3$.

The integrand $j(\xi',\phi,e)$ in the integral over the history of charge motion $J(\bar{\xi},\phi,e)$ expands in powers of orbital eccentricity up to the second order as
\begin{align}
j^{(2)}(\xi',\phi,e) & = i^{(0)}(\xi',\phi) + e \, i^{(1)}(\xi',\phi) + e^2 \, i^{(2)}(\xi',\phi) + {\cal O}(e^3), \\
i^{(2)}(\xi',\phi) & = - \frac{\sin \xi' \sin \phi + 4 \cos \xi' \cos \phi - 6 \cos^2 \xi' ( \sin \xi' \sin \phi + \cos \xi' \cos \phi)}{2\sqrt{\bar{\xi} - \xi'}} \nonumber \\
& + \frac{(\sin \bar{\xi} - \sin \xi') (2 \cos \xi' (\sin \xi' \sin \phi + \cos \xi' \cos \phi) - \cos \phi)}{2 (\bar{\xi} - \xi')^{3/2}} \nonumber \\
& + \frac{3 ( \sin \bar{\xi} - \sin \xi' )^2 ( \sin \xi' \sin \phi + \cos \xi' \cos \phi )}{8 (\bar{\xi} - \xi')^{5/2}}.
\end{align}
Here terms $i^{(0)}(\xi',\phi)$ and $i^{(1)}(\xi',\phi)$ are determined by the Eqs. \eqref{eq:i^(0)_def} and \eqref{eq:i^(1)_def}. Accordingly, integral over the history of charge motion splits into three integrals
\begin{align}
\label{eq:J^(2)_decomp}
& J^{(2)}(\bar{\xi},\phi,e) = \int_{-\infty}^{\bar{\xi}} d\xi' \, j^{(2)}(\xi',\phi,e) = I^{(0)}(\bar{\xi},\phi) + e I^{(1)}(\bar{\xi},\phi) + e^2 I^{(2)}(\bar{\xi},\phi) + {\cal O}(e^3), \\
& I^{(2)}(\bar{\xi},\phi) = \int_{-\infty}^{\bar{\xi}} d\xi' \, i^{(2)}(\xi',\phi).
\end{align}
Integrals $I^{(0)}(\bar{\xi},\phi)$ and $I^{(1)}(\bar{\xi},\phi)$ are computed in the previous section and given by the Eqs. \eqref{eq:I^(0)_exp} and \eqref{eq:I^(1)_exp}. Thus, we have to calculate only the integral $I^{(2)}(\bar{\xi},\phi)$.

By analogy with the computation of $I^{(1)}(\bar{\xi},\phi)$, we split $i^{(2)}(\xi',\phi)$ into three parts with different inverse powers of $\sqrt{\bar{\xi} - \xi'}$
\begin{align}
& i^{(2)}(\xi',\phi) = i^{(2,1)}(\xi',\phi) + i^{(2,3)}(\xi',\phi) + i^{(2,5)}(\xi',\phi), \\
& i^{(2,1)}(\xi',\phi) = - \frac{\sin \xi' \sin \phi + 4 \cos \xi' \cos \phi - 6 \cos^2 \xi' ( \sin \xi' \sin \phi + \cos \xi' \cos \phi)}{2\sqrt{\bar{\xi} - \xi'}}, \\
& i^{(2,3)}(\xi',\phi) = \frac{(\sin \bar{\xi} - \sin \xi') (2 \cos \xi' (\sin \xi' \sin \phi + \cos \xi' \cos \phi) - \cos \phi)}{2 (\bar{\xi} - \xi')^{3/2}}, \\
& i^{(2,5)}(\xi',\phi) = \frac{3 ( \sin \bar{\xi} - \sin \xi' )^2 ( \sin \xi' \sin \phi + \cos \xi' \cos \phi )}{8 (\bar{\xi} - \xi')^{5/2}}.
\end{align}
Correspondingly, the integral $I^{(2)}(\bar{\xi},\phi)$ splits into three integrals
\begin{align}
& I^{(2)}(\bar{\xi},\phi) = I^{(2,1)}(\bar{\xi},\phi) + I^{(2,3)}(\bar{\xi},\phi) + I^{(2,5)}(\bar{\xi},\phi), \\
& I^{(2,1)}(\bar{\xi},\phi) = \int_{-\infty}^{\bar{\xi}} d\xi' \, i^{(2,1)}(\xi',\phi), \quad I^{(2,3)}(\bar{\xi},\phi) = \int_{-\infty}^{\bar{\xi}} d\xi' \, i^{(2,3)}(\xi',\phi), \\
& I^{(2,5)}(\bar{\xi},\phi) = \int_{-\infty}^{\bar{\xi}} d\xi' \, i^{(2,5)}(\xi',\phi).
\end{align}
The obtained integrals are computed by the common scheme -- integrating by parts we reduce them to the combination of the Fresnel integrals \eqref{eq:Fres_int}.

We start with the calculation of $I^{(2,1)}(\bar{\xi},\phi)$. Transforming the integration variable as $s = \bar{\xi} - \xi'$ and using the Eq. \eqref{eq:Fres_int}, we arrive at
\begin{equation}
\label{eq:I^(2,1)_exp}
I^{(2,1)}(\bar{\xi},\phi) = \frac{1}{4} \sqrt{\pi} \left \lbrack \sin \left( \bar{\xi} - \phi + \frac{\pi}{4} \right) + \sqrt{3} \cos \left( \phi + \frac{\pi}{4} - 3\bar{\xi} \right) \right \rbrack.
\end{equation}

To calculate the integral $I^{(2,3)}(\bar{\xi},\phi)$, we integrate it by parts once
\begin{align}
I^{(2,3)}(\bar{\xi},\phi) & = \int_{-\infty}^{\bar{\xi}} d\xi' \, \frac{(\sin \bar{\xi} - \sin \xi') (2 \cos \xi' (\sin \xi' \sin \phi + \cos \xi' \cos \phi) - \cos \phi)}{2 (\bar{\xi} - \xi')^{3/2}} \nonumber \\ & = \sqrt{\varepsilon} \cos \bar{\xi} \Big( 2 \cos \bar{\xi} (\sin \bar{\xi} \sin \phi + \cos \bar{\xi} \cos \phi) - \cos \phi \Big) + \int _{-\infty}^{\bar{\xi}} \frac{d\xi'}{\sqrt{\bar{\xi} - \xi'}} \nonumber \\ & \times \Big \lbrack \cos \xi' \Big( 2 \cos \xi' ( \cos \xi' \cos \phi + \sin \xi' \sin \phi ) - \cos \phi \Big) - ( \sin \bar{\xi} - \sin \xi' ) \nonumber \\ & \times \Big( 2 \cos \xi' ( \cos \xi' \sin \phi - \cos \phi \sin \xi' ) - 2 \sin \xi' ( \cos \xi' \cos \phi + \sin \xi' \sin \phi ) \Big) \Big \rbrack.
\end{align}
Here, in the first line, we implicitly introduced the regularizing parameter $\bar{\xi} - \varepsilon$, $\varepsilon \to +0$ into the upper integration limit. The boundary term resulting from the integration by parts vanishes. Changing the integration variable $s = \bar{\xi} - \xi'$, by use of the Eq. \eqref{eq:Fres_int} we find
\begin{equation}
\label{eq:I^(2,3)_exp}
I^{(2,3)}(\bar{\xi},\phi) = \frac{1}{2} \sqrt{\pi} \left \lbrack \left( \sqrt{2} - 1 \right) \sin \left( \bar{\xi} - \phi + \frac{\pi}{4} \right) + \left( \sqrt{3} - \sqrt{2} \right) \cos \left( - 3\bar{\xi} + \phi + \frac{\pi}{4} \right) \right \rbrack.
\end{equation}

Computing the integral $I^{(2,5)}(\bar{\xi},\phi)$ we have to integrate it by parts twice. After the first integration by parts it takes the form
\begin{align}
I^{(2,5)}(\bar{\xi},\phi) & = \frac{3 ( \sin \bar{\xi} - \sin \xi' )^2 ( \sin \xi' \sin \phi + \cos \xi' \cos \phi )}{8 (\bar{\xi} - \xi')^{5/2}} \nonumber \\ & = \frac{1}{4} \sqrt{\varepsilon} \cos^2 \bar{\xi} ( \sin \bar{\xi} \sin \phi + \cos \bar{\xi} \cos \phi ) - \int_{-\infty}^{\bar{\xi}} d\xi' \, \frac{(\sin \bar{\xi} - \sin \xi')}{4(\bar{\xi} - \xi')^{3/2}} \Big \lbrack (\sin \bar{\xi} - \sin \xi') \nonumber \\ & \times ( \cos \xi' \sin \phi - \cos \phi \sin \xi' ) - 2 \cos \xi' (\cos \xi' \cos \phi + \sin \xi' \sin \phi) \Big \rbrack.
\end{align}
After the second integration by parts $I^{(2,5)}(\bar{\xi},\phi)$ is written as
\begin{align}
I^{(2,5)}(\bar{\xi},\phi) & = - \frac{1}{2} \varepsilon^{3/2} \cos^2 \bar{\xi} (\cos \bar{\xi} \sin \phi - \sin \bar{\xi} \cos \phi) + \int_{-\infty}^{\bar{\xi}} \frac{d\xi'}{2\sqrt{\bar{\xi} - \xi'}} \nonumber \\ & \times \Big \lbrack 2 \cos^2 \xi' (\cos \xi' \cos 
\phi + \sin \xi' \sin \phi) - 4 \cos \xi' (\sin \bar{\xi} - \sin \xi') \nonumber \\ & \times (\cos \xi' \sin \phi - \cos \phi \sin \xi') - (\cos \xi' \cos \phi + \sin \xi' \sin \phi) (\sin \bar{\xi} - \sin \xi')^2 \nonumber \\ & + 2 \sin \xi' (\cos \xi' \cos \phi + \sin \xi' \sin \phi) (\sin \bar{\xi} - \sin \xi') \Big \rbrack.
\end{align}
Note that, by analogy with the previous integrals, here the boundary terms resulting from the integration by parts vanish. Therefore, after the change of integration variable $s = \bar{\xi} - \xi'$ and some algebra the integral $I^{(2,5)}(\bar{\xi},\phi)$ reduces to the combination of Fresnel integrals \eqref{eq:Fres_int}. Finally, we find it in the form
\begin{align}
\label{eq:I^(2,5)_exp}
I^{(2,5)}(\bar{\xi},\phi) & = \frac{1}{8} \sqrt{\pi} \left \lbrack \sin \left( \bar{\xi} + \phi + \frac{\pi}{4} \right) + 4(\sqrt{2} - 1) \sin \left( \bar{\xi} - \phi + \frac{\pi}{4} \right) + \cos \left( \bar{\xi} + \phi + \frac{\pi}{4} \right) \right. \nonumber \\ & \left. - ( 4\sqrt{2} - 3\sqrt{3} - 1 ) \cos \left( - 3\bar{\xi} + \phi + \frac{\pi}{4} \right) \right \rbrack.
\end{align}

\begin{figure}[t]
\center{\includegraphics[width=0.65\linewidth]{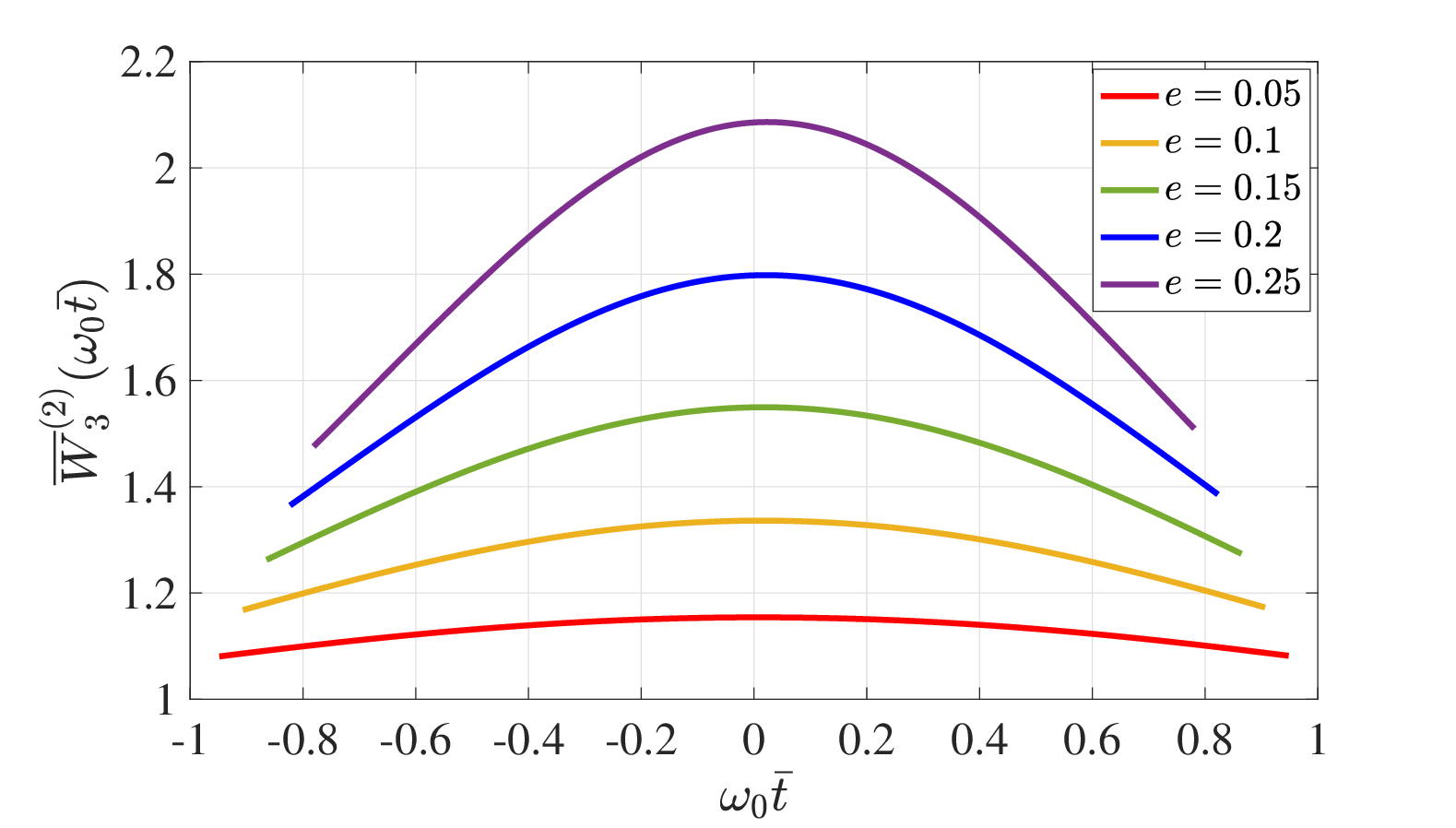}}
\caption{Normalised radiation power of the charge on elliptical orbit in three spacetime dimensions in the quadratic approximation in eccentricity for $e = 0.05, \, 0.1, \, 0.15, \, 0.2, \, 0.25$. The tails in radiation manifest distinctly with increasing orbital eccentricity.}
\label{fig:4}
\end{figure}

Combining the Eqs. \eqref{eq:I^(2,1)_exp}, \eqref{eq:I^(2,3)_exp} and \eqref{eq:I^(2,5)_exp}, we obtain the following expression for the integral $I^{(2)}(\bar{\xi},\phi)$
\begin{align}
\label{eq:I^(2)_exp}
I^{(2)}(\bar{\xi},\phi) & = \frac{1}{8} \sqrt{\pi} \left \lbrack - 6\sin \left( \bar{\xi} - \phi + \frac{\pi}{4} \right) + \sin \left( \bar{\xi} + \phi + \frac{\pi}{4} \right) + 8\sqrt{2} \sin \left( \bar{\xi} - \phi + \frac{\pi}{4} \right) \right. \nonumber \\ & + \left. \cos \left( \bar{\xi} + \phi + \frac{\pi}{4} \right) - (8\sqrt{2} - 9\sqrt{3} - 1) \cos \left( - 3\bar{\xi} + \phi + \frac{\pi}{4} \right) \right \rbrack.
\end{align}
Thus, using the Eqs. \eqref{eq:I^(0)_exp}, \eqref{eq:I^(1)_exp} and \eqref{eq:J^(2)_decomp}, we find the integral over the history of charge motion up to the contributions quadratic in the orbital eccentricity as
\begin{align}
\label{eq:hist_int_ellipt_2nd}
J^{(2)}(\bar{\xi},\phi,e) & = \frac{1}{8} \sqrt{\pi} \left \lbrack 4\sqrt{2} \cos (\phi - \bar{\xi}) - 4\sqrt{2} \sin (\phi - \bar{\xi}) + 2\sqrt{2}e \cos \phi - 2\sqrt{2}e \sin \phi \right. \nonumber \\ & + \left. (8 - 2\sqrt{2}) e \cos (\phi - 2\bar{\xi}) - (8 - 2\sqrt{2}) e \sin (\phi - 2\bar{\xi}) + e^2 \sin \left( \bar{\xi} + \phi + \frac{\pi}{4} \right) \right. \nonumber \\ & \left. + e^2 \cos \left( \bar{\xi} + \phi + \frac{\pi}{4} \right) + (8\sqrt{2} - 6) e^2 \sin \left( \bar{\xi} - \phi + \frac{\pi}{4} \right) \right. \nonumber \\ &  \left. - (8\sqrt{2} - 9\sqrt{3} - 1)e^2 \cos \left( 3\bar{\xi} - \phi - \frac{\pi}{4} \right) \right \rbrack.
\end{align}
As a result, integrating the square of $J^{(2)}(\bar{\xi},\phi,e)$ over the polar angle $\phi$, we obtain the following expression for the radiation power of a non-relativistic charge on elliptical orbit in three spacetime dimensions
\begin{align}
\label{eq:Rad_pow_2nd}
W_{3}^{(2)}(\bar{\xi}) & = \frac{\pi}{128} g^2 \omega_0^3 a^2 \Big \lbrack 32 + 64\sqrt{2}e \cos \bar{\xi} + 32(1 + \sqrt{2})e^2 + 8 (9\sqrt{3} - 4\sqrt{2}) e^2 \cos 2\bar{\xi} \nonumber \\ & + 8e^2 \sin 2\bar{\xi} + 4e^3 \sin \bar{\xi} - 4( 2\sqrt{2} + 9\sqrt{3} - 18\sqrt{6})e^3 \cos \bar{\xi} + 4(2\sqrt{2} - 1)e^3 \sin 3\bar{\xi} \nonumber \\ & + 12(3\sqrt{3} - 2\sqrt{2})e^3 \cos 3\bar{\xi} + (269 + 9\sqrt{3} - 56\sqrt{2} - 72\sqrt{6})e^4 + (8\sqrt{2} - 6)e^4 \sin 2\bar{\xi} \nonumber \\ & + 2( 32\sqrt{2} - 27\sqrt{3} + 36\sqrt{6} - 70)e^4 \cos 2\bar{\xi} + (9\sqrt{3} + 1 - 8\sqrt{2})e^4 \sin 4\bar{\xi} \nonumber \\ & + (9\sqrt{3} + 1 - 8\sqrt{2})e^4 \cos 4\bar{\xi} \Big \rbrack.
\end{align}
By analogy with the previous section, we call it the quadratic approximation of the radiation power. Normalising the radiation power to its value for a circular orbit
\begin{equation}
\label{eq:Rad_pow_2nd_norm}
\overline{W}_{3}^{(2)}(\bar{\xi}) = \frac{4}{\pi g^2 \omega_0^3 a^2} W_{3}^{(2)}(\bar{\xi}),
\end{equation}
by use of the Eq. \eqref{eq:time_anom_rel} we obtain the dependence of radiation power on the retarded time for various values of the eccentricity $e \leq 0.3$ (see Fig. \eqref{fig:4}). From Fig. \eqref{fig:4} one finds that the radiation power takes its maximum value after the charge passes through the pericenter of the orbit, which corresponds to the time $\bar{t}=0$. This is the distinct signature of the tail term in radiation of the charge on elliptical orbit.

\begin{figure}[t]
\begin{minipage}[h]{0.49\linewidth}
\center{\includegraphics[width=0.95\linewidth]{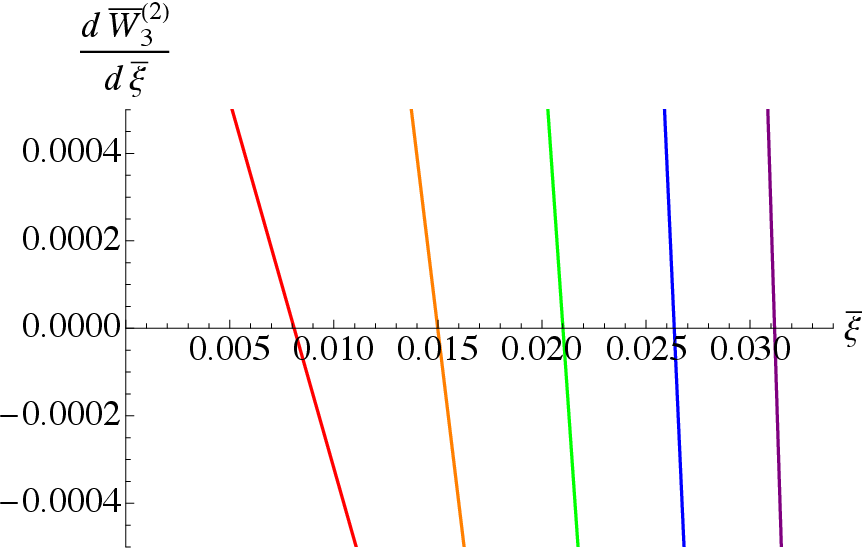}}
\end{minipage}
\hfill
\begin{minipage}[h]{0.49\linewidth}
\center{\includegraphics[width=0.95\linewidth]{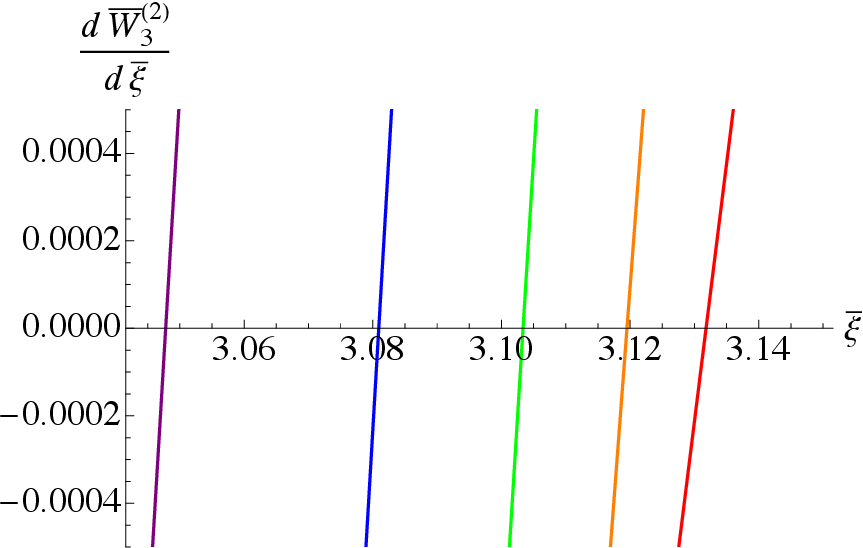}}
\end{minipage}
\caption{The first derivative of normalised radiation power of the charge on elliptical orbit for $e = 0.05, \, 0.1, \, 0.15, \, 0.2, \, 0.25$. The radiation power extremum points are shifted in time from the moments when the charge passes through the pericenter and apocenter of the orbit (here $\bar{\xi}=0$ and $\bar{\xi} = \pi$ correspondingly). The colors denote the same eccentricity values as on the Fig. \eqref{fig:4}.}
\label{fig:5}
\end{figure}

Let us find the dependence of the radiation power extremum points on the orbital eccentricity
\begin{align}
\label{eq:Rad_pow_2nd_norm_der}
\frac{d \overline{W}_{3}^{(2)}}{d\bar{\xi}} & = \frac{1}{32} \Big \lbrack - 64\sqrt{2}e \sin \bar{\xi} + 16e^2 \cos 2\bar{\xi} - 16(9\sqrt{3} - 4\sqrt{2})e^2 \sin 2\bar{\xi} + 4e^3 \cos \bar{\xi} \nonumber \\ & + 4( 2\sqrt{2} + 9\sqrt{3} - 18\sqrt{6})e^3 \sin \bar{\xi} + 12( 2\sqrt{2} - 1)e^3 \cos 3\bar{\xi} - 36(3\sqrt{3} - 2\sqrt{2}) \nonumber \\ & \times e^3 \sin 3\bar{\xi} + 4( 70 - 32\sqrt{2} + 27\sqrt{3} - 36\sqrt{6})e^4 \sin 2\bar{\xi} + 4(4\sqrt{2} - 3)e^4 \cos 2\bar{\xi} \nonumber \\ & - 4(9\sqrt{3} + 1 - 8\sqrt{2})e^4 \sin 4\bar{\xi} + 4(9\sqrt{3} + 1 - 8\sqrt{2})e^4 \cos 4\bar{\xi} \Big \rbrack = 0.
\end{align}
First, we find the roots of Eq. \eqref{eq:Rad_pow_2nd_norm_der} graphically. From Fig. \eqref{fig:5} one finds that the radiation power takes its maximum values after the charge passes through the pericenter of the orbit corresponding to $\bar{\xi} = 2\pi n$, $n \in \mathbb{Z}$. Moreover, the higher the eccentricity of the orbit, the greater the shift of the extremum points. Interestingly, the minimum points of the radiation power are shifted in the opposite direction -- it takes its minimum values before the charge passes through the apocenter of the orbit $\bar{\xi} = (2n+1)\pi$, $n \in \mathbb{Z}$.

Let us find the extremum points of the radiation power \eqref{eq:Rad_pow_2nd} up to the contributions quadratic in the orbital eccentricity. After the rearrangement of some terms, Eq. \eqref{eq:Rad_pow_2nd_norm_der} is written as
\begin{multline}
\label{eq:extr_eq_e^2}
- \sin \bar{\xi} + \frac{e}{4\sqrt{2}} \left \lbrack \cos 2\bar{\xi} - (9\sqrt{3} - 4\sqrt{2}) \sin 2\bar{\xi} \right \rbrack + \frac{e^2}{16\sqrt{2}} \left \lbrack \cos \bar{\xi} + ( 2\sqrt{2} + 9\sqrt{3} - 18\sqrt{6}) \sin \bar{\xi} \right \rbrack \\ + \frac{3e^2}{16\sqrt{2}} \left \lbrack ( 2\sqrt{2} - 1) \cos 3\bar{\xi} - 3 (3\sqrt{3} - 2\sqrt{2}) \sin 3\bar{\xi} \right \rbrack + {\cal O}(e^3) = 0.
\end{multline}
We rewrite the expressions in square brackets as a sine of sum or difference. The first bracket in the Eq. \eqref{eq:extr_eq_e^2} takes the form
\begin{align}
& \cos 2\bar{\xi} - (9\sqrt{3} - 4\sqrt{2}) \sin 2\bar{\xi} = -\frac{1}{N_1} \sin(2\bar{\xi}-\alpha), \\
& \sin \alpha = N_1, \quad \cos \alpha = N_1 (9\sqrt{3} - 4\sqrt{2}), \quad N_1=\left( 1 + (9\sqrt{3}-4\sqrt{2})^2 \right)^{-1/2}.
\end{align}
Analogously, the second bracket in the Eq. \eqref{eq:extr_eq_e^2} is written as
\begin{align}
& \cos \bar{\xi} + ( 2\sqrt{2} + 9\sqrt{3} - 18\sqrt{6}) \sin \bar{\xi} = \frac{1}{N_2} \sin(\bar{\xi}+\beta), \\
& \sin \beta = N_2, \quad \cos \beta = N_2 (2\sqrt{2}+9\sqrt{3}-18\sqrt{6}), \\
& N_2 = \left( 1 + (2\sqrt{2}+9\sqrt{3}-18\sqrt{6})^2 \right)^{-1/2}.
\end{align}
Finally, rewriting the third bracket in the Eq. \eqref{eq:extr_eq_e^2} we arrive at
\begin{align}
& ( 2\sqrt{2} - 1) \cos 3\bar{\xi} - 3 (3\sqrt{3} - 2\sqrt{2}) \sin 3\bar{\xi} = - \frac{1}{N_3} \sin(3\bar{\xi}-\delta), \\
& \sin \delta = N_3 (2\sqrt{2}-1), \quad \cos \delta = 3 N_3 (3\sqrt{3}-2\sqrt{2}), \\
& N_3 = \left( (2\sqrt{2}-1)^2 + 9(3\sqrt{3}-2\sqrt{2})^2 \right)^{-1/2}.
\end{align}
As a result, the Eq. \eqref{eq:extr_eq_e^2} takes the form
\begin{align}
\label{eq:extr_eq_trig_form}
& - \sin \bar{\xi} - e_1 \sin(2\bar{\xi}-\alpha) + e_{2}^{2} \sin{\bar{\xi}+\beta} - e_{3}^{2} \sin(3\bar{\xi}-\delta) = 0, \\
& e_1 = \frac{e}{4\sqrt{2}N_1}, \quad e_{2}^{2} = \frac{e^2}{16\sqrt{2}N_2}, \quad e_{3}^{2} = \frac{3e^2}{16\sqrt{2}N_3}.
\end{align}
From Fig. \eqref{fig:5} one finds that the radiation power extremum points are close to the points $\bar{\xi} = \pi n, \, n \in \mathbb{Z}$. Thus, we can represent the roots of the Eq. \eqref{eq:extr_eq_trig_form} as
\begin{equation}
\bar{\xi} = \pi n + \Delta \bar{\xi}, \quad n \in \mathbb{Z},
\end{equation}
where $\Delta \bar{\xi} \ll 1$ are the small shifts of the extremum points from the moments when the charge passes through the pericenter and apocenter of the orbit. Expanding the Eq. \eqref{eq:extr_eq_trig_form} up to the leading contribution in $\Delta \bar{\xi}$, we find the radiation power extremum points in the following form
\begin{equation}
\label{eq:extr_pos}
\bar{\xi} = \pi n + (-1)^n \frac{e_1 \alpha + (-1)^n e_{2}^{2} \beta + (-1)^n e_{3}^{2} \delta}{1 + 2(-1)^n e_1 - e_{2}^2 + 3 e_{3}^{2}}.
\end{equation}
Here the following approximations can be used with sufficient accuracy
\begin{equation}
\alpha \simeq N_1, \quad \beta \simeq N_2, \quad \delta \simeq (2\sqrt{2}-1)N_3.
\end{equation}
In accordance with Fig. \eqref{fig:5}, from the Eq. \eqref{eq:extr_pos} one finds that the maximums of radiation power are shifted forward in time with respect to the moments when the charge passes through the pericenter of the orbit $2\pi n, \, n \in \mathbb{Z}$
\begin{equation}
\label{eq:max_shifts}
\Delta \bar{\xi}_{\rm max} = \frac{e_1\alpha + e_{2}^{2} \beta + e_{3}^2 \delta}{1 + 2 e_1 - e_{2}^2 + 3 e_{3}^2} > 0.
\end{equation}
The shift value increases quadratically with the orbital eccentricity. Analogously, the minimums of radiation power are shifted backward in time with respect to the moments when the charge passes through the apocenter of the orbit $(2n+1)\pi, \, n \in \mathbb{Z}$
\begin{equation}
\label{eq:min_shifts}
\Delta \bar{\xi}_{\rm min} = - \frac{e_1\alpha - e_{2}^2 \beta - e_{3}^2 \delta}{1 - 2 e_1 - e_{2}^2 + 3 e_{3}^2} < 0,
\end{equation}
in accordance with Fig. \eqref{fig:5}. The obtained formulae \eqref{eq:max_shifts} and \eqref{eq:min_shifts} for the shifts of the radiation power extremum points are in agreement with the graphical solutions to the Eq. \eqref{eq:Rad_pow_2nd_norm_der} for the eccentricity in range $e \in (0, 0.3)$.

\section{Spectral distribution of radiation}\label{IV}

Let us also compare the spectral distributions of radiation of the charge on elliptical orbit in three and four spacetime dimensions.

We use the formula obtained in Ref. \cite{Galtsov:2020hhn} for the spectral-angular distribution of the total energy of scalar radiation of an arbitrarily moving charge in $D=n+1$ dimensions
\begin{align}
& \frac{dE_{n+1}}{d\omega d\Omega_{n-1}} = \frac{\Omega_{n-1} \omega^{n-1}}{2(2\pi)^n} \left. | \tilde{j}(p) |^2 \right \vert_{p^0 = |\mathbf{p}| = \omega}, \\
& \tilde{j}(p) = g \int d\tau \exp \left \lbrack i \omega t(\tau) - i \mathbf{p} \mathbf{z}(\tau) \right \rbrack,
\end{align}
where $t(\tau)=z^0(\tau)$ is the time coordinate of the charge. In the Fourier image of the current, we transform the integral over proper time to the integral over coordinate time
\begin{align}
d\tau = \sqrt{1-\mathbf{v}^2} dt \quad \Longrightarrow \quad \tilde{j}(p) = g \int dt \, \sqrt{1-\mathbf{v}^2} \exp \left \lbrack i \omega t - i \mathbf{p} \mathbf{z}(t) \right \rbrack,
\end{align}
where $\mathbf{v} = d\mathbf{z}/dt$. As a result, the spectral-angular distribution of the total radiation energy takes the form
\begin{equation}
\frac{dE_{n+1}}{d\omega d\Omega_{n-1}} = \frac{\Omega_{n-1} \omega^{n-1} g^2}{2(2\pi)^n} \left \vert \int dt \, \sqrt{1-\mathbf{v}^2} \exp \left \lbrack i \omega t - i \mathbf{p} \mathbf{z}(t) \right \rbrack \right \vert^2.
\end{equation}

In the case of periodic motion of the charge, the integral over time can be represented as
\begin{equation}
\int dt \, \sqrt{1-\mathbf{v}^2} \exp \left \lbrack i \omega t - i \mathbf{p} \mathbf{z}(t) \right \rbrack = \sum_{l=-\infty}^{\infty} e^{i l \omega T} \int_{0}^{T} dt \, \sqrt{1-\mathbf{v}^2} \exp \left \lbrack i \omega t - i \omega \mathbf{n} \mathbf{z}(t) \right \rbrack,
\end{equation}
where $\mathbf{p} = \omega \mathbf{n}$ and $T$ is the period of charge motion. Using the relation \cite{Borodovitsyn_book}
\begin{equation}
\sum_{l=-\infty}^{\infty} e^{i l x} = 2\pi \sum_{l=-\infty}^{\infty} \delta(x - 2 \pi l)
\end{equation}
we rewrite the integral over time as
\begin{equation}
\omega_0 \sum_{l=-\infty}^{\infty} \delta(\omega - \omega_0 l)  I_{n+1}^{l}, \quad I_{n+1}^{l} = \int_0^T dt \, \sqrt{1-\mathbf{v}^2} \exp \left \lbrack i \omega_0 l (t - \mathbf{n}\mathbf{z}) \right \rbrack.
\end{equation}
where $\omega_0=2\pi/T$ is a frequency of orbital motion of the charge. Accordingly, the square of its absolute value is written as
\begin{equation}
\left \vert \int dt \, \sqrt{1-\mathbf{v}^2} \exp \left \lbrack i \omega t - i \mathbf{p} \mathbf{z}(t) \right \rbrack \right \vert^2 = \omega_0^2 \delta(0) \sum_{l=-\infty}^{\infty} \delta(\omega - \omega_0 l) |I_{n+1}^{l}|^2.
\end{equation}
Here we consider $\delta(0)$ as the following limit
\begin{equation}
2\pi \delta(0) = \lim_{T_0 \to \infty} \lim_{\omega \to 0} \int_{-T_0/2}^{T_0/2} dt \, e^{i \omega t} = \lim_{T_0 \to \infty} T_0,
\end{equation}
where $T_0$ is the total time of charge motion. Thus, defining the radiation power as
\begin{equation}
W_{n+1} = \lim_{T_0 \to \infty} \frac{E_{n+1}}{T_0},
\end{equation}
we find the spectral-angular distribution of the charge radiation power in the form
\begin{equation}
\label{eq:n+1_sp_ang_distr}
\frac{dW_{n+1}}{d\omega d\Omega_{n-1}} = \frac{\Omega_{n-1} \omega^{n-1} g^2}{2(2\pi)^{n+1}} \omega_0^2 \sum_{l=-\infty}^{\infty} \delta(\omega - \omega_0 l) |I_{n+1}^{l}|^2.
\end{equation}
As one could expect, in $D=n+1$ dimensions the radiation spectrum of a periodically moving charge is discrete with frequencies being multiples of the orbital frequency. Finally, integrating the Eq. \eqref{eq:n+1_sp_ang_distr}, we obtain the spectral distribution of radiation power of an arbitrarily moving charge in $D=n+1$ dimensions
\begin{equation}
\label{eq:n+1_arbit_spectrum}
W_{n+1} = \sum_{l=1}^{\infty} \frac{\Omega_{n-1} \omega_0^{n+1} l^{n-1} g^2}{2(2\pi)^{n+1}} \int d\Omega_{n-1} \, |I_{n+1}^{l}|^2 \equiv \sum_{l=1}^{\infty} W_{n+1}^{l},
\end{equation}
where $W_{n+1}^{l}$ is the radiation power on the $l$-th harmonic of the spectrum.

For the charge on elliptical orbit, the integral $I_{n+1}^{l}$ can be calculated simultaneously for three and four spacetime dimensions. Note that if in four dimensions the charge moves inside the equatorial plane $\mathbf{z}(t) = \lbrace \rho(t) \cos \psi(t), \rho(t) \sin \psi(t), 0 \rbrace$, we find
\begin{equation}
\mathbf{n} \mathbf{z}(t) = \rho(t) \cos \psi(t) \cos \phi \sin \theta + \rho(t) \sin \psi(t) \sin \phi \sin \theta.
\end{equation}
In turn, in three dimensions from the Eq. \eqref{eq:2D_ellipt_WL} we obtain
\begin{equation}
\mathbf{n} \mathbf{z}(t) = \rho(t) \cos \psi(t) \cos \phi + \rho(t) \sin \psi(t) \sin \phi.
\end{equation}
Thus, the integral $I_3^l$ can be found from the integral $I_4^l$ by the following substitution
\begin{equation}
I_3^l = \left. I_4^l \right \vert_{\theta = \pi/2}.
\end{equation}
Therefore, we have to calculate only the integral $I_4^l$ to find the spectral distributions in three and four dimensions.

It is convenient to pass in $I_4^l$ from the integral over time to the integral over anomaly $\xi$ by use of the Eqs. \eqref{eq:t-anom_rel} and \eqref{eq:WL_of_anom}. Also, using these relations we find the charge velocity as a function of anomaly
\begin{equation}
\mathbf{v}^2(\xi) = a^2 \omega_0^2 \frac{1 + e \cos \xi}{1 - e \cos \xi}.
\end{equation}
As a result, the integral $I_4^l$ takes the following form
\begin{align}
\label{eq:I_4^l_gen}
I_4^l & = \frac{1}{\omega_0} \int_{0}^{2\pi} d\xi \, (1 - e \cos \xi) \sqrt{1 - a^2 \omega_0^2 \frac{1 + e \cos \xi}{1 - e \cos \xi}} \exp \big \lbrack i l \big( \xi - e \sin \xi \nonumber \\ & - \omega_0 a (\cos \xi - e) \cos \phi \sin \theta - \omega_0 b \sin \xi \sin \phi \sin \theta \big) \big \rbrack,
\end{align}
where $b=a\sqrt{1-e^2}$ is a minor semiaxis of the ellipse. The Eq. \eqref{eq:I_4^l_gen} determines the spectral distribution of radiation of the relativistic charge moving along a fixed elliptical orbit with an arbitrary velocity. In general, calculation of this integral is complicated. We calculate it in two limiting cases -- for the relativistic charge on a circular orbit and for the non-relativistic charge on an elliptical orbit. The characteristic differences between the spectral distributions of radiation in three and four dimensions already manifest themselves in these two simple cases.

\subsection{Charge on a circular orbit}

For the charge on a circular orbit $e=0$, the integral \eqref{eq:I_4^l_gen} can be calculated exactly for an arbitrary value of the charge velocity. Here we have
\begin{equation}
e=0 \quad \Longrightarrow \quad a=b=R_0, \quad v_0 \equiv |\mathbf{v}| = R_0 \omega_0 = {\rm Const},
\end{equation}
where $R_0$ is the radius of circular orbit and $v_0$ is the constant charge velocity. As a result, the integral $I_4^l$ is simplified significantly
\begin{equation}
I_4^l = \frac{1}{\gamma \omega_0} \int_{0}^{2\pi} d\xi \, \exp \left \lbrack i l \left( \xi - v_0 \sin \left(\xi - \phi + \frac{\pi}{2} \right) \sin \theta \right) \right \rbrack,
\end{equation}
where $\gamma = (1-v_0^2)^{-1/2}$ is the relativistic Lorentz-factor of the charge. By analogy with the Ref. \cite{Galtsov:2020hhn}, using relations
\begin{equation}
e^{-i x \sin a} = \sum_{\nu = - \infty}^{\infty} J_\nu(x) e^{-i \nu a}, \quad \int_{0}^{2\pi} d\xi \, e^{i(l-\nu)\xi} = 2\pi \delta_{l\nu},
\end{equation}
we calculate the integral $I_4^l$ arriving at
\begin{equation}
\label{eq:4D_circ_I^l_int}
I_4^l = \frac{2\pi}{\gamma \omega_0} e^{i l (\phi - \pi/2)} J_l(v_0l \sin \theta).
\end{equation}
By use of this, we find the spectral distributions of radiation in three and four dimensions.

In three spacetime dimensions, we find the integral $I_3^l$ as
\begin{equation}
I_3^l = \left. I_4^l \right \vert_{\theta = \pi/2} = \frac{2\pi}{\gamma \omega_0} e^{i l (\phi - \pi/2)} J_l(v_0 l).
\end{equation}
Thus, substituting it into the Eq. \eqref{eq:n+1_arbit_spectrum} and integrating over the polar angle $\phi$ we find the spectral distribution of radiation power in the form
\begin{equation}
\label{eq:2+1_circ_spectrum}
W_3 = \frac{\pi \omega_0 g^2}{\gamma^2} \sum_{l=1}^{\infty} l J_l^2(v_0 l).
\end{equation}
For convenience of comparing this with its four-dimensional analog, we introduce the normalised spectral distribution of radiation power of the charge on a circular orbit in three dimensions
\begin{equation}
\label{eq:3D_circ_spect_distr}
\overline{W}_{n+1}^{l} = \frac{W_{n+1}^{l}}{W_{n+1}} \quad \Longrightarrow \quad \overline{W}_3^l = \frac{l J_l^2(v_0 l)}{\sum_{n=1}^{\infty} n J_n^2(v_0 n)}.
\end{equation}
In what follows, we numerically estimate the obtained three-dimensional normalising factor in the Eq. \eqref{eq:3D_circ_spect_distr}.

Now we turn to the calculation of the four-dimensional spectral distribution of radiation power. From the Eqs. \eqref{eq:n+1_arbit_spectrum} and \eqref{eq:4D_circ_I^l_int} we find that it is determined by the following series
\begin{equation}
\label{eq:e=0_D=4_distr_start}
W_4 = \sum_{l=1}^{\infty} \frac{\omega_0^2 l^2 g^2}{\gamma^2} \int d\theta \, \sin \theta \, J_l^2(v_0 l \sin \theta).
\end{equation}
By use of two subsequent changes of the integration variable
\begin{equation}
x = \cos \theta \quad \Longrightarrow \quad y=\sqrt{1-x^2}
\end{equation}
we compute the remaining integral as \cite{Zwillinger_table}
\begin{equation}
\int d\theta \, \sin \theta \, J_l^2(v_0 l \sin \theta) = 2 \frac{(v_0 l)^{2l}}{(2l+1)!} {_1}F_2 \left( l + \frac{1}{2}; l + \frac{3}{2}, 1 + 2l; - l^2 v_0^2 \right),
\end{equation}
where ${_1}F_2(a;b_1,b_2;z)$ is the generalised hypergeometric function. Thus, the spectral distribution of radiation power in four spacetime dimensions takes the form
\begin{equation}
W_4 = 2 \frac{\omega_0^2 g^2}{\gamma^2} \sum_{l=1}^{\infty} \frac{l^{2(l+1)}}{(2l+1)!} v_0^{2l} {_1}F_2 \left( l + \frac{1}{2}; l + \frac{3}{2}, 1 + 2l; - l^2 v_0^2 \right).
\end{equation}
To construct the normalised distribution, we have to sum this spectrum. We return to the Eq. \eqref{eq:e=0_D=4_distr_start} and sum the series before calculating the angular integral \cite{Borodovitsyn_book}
\begin{equation}
\sum_{l=1}^{\infty} l^2 J_l^2(v_0 l \sin \theta) = \frac{v_0^2 \sin^2 \theta (4 + v_0^2 \sin^2 \theta)}{2^4(1 - v_0^2 \sin^2 \theta)^{7/2}}.
\end{equation}
The resulting angular integral can be easily calculated \cite{Borodovitsyn_book}. As a result, the four-dimensional radiation power of the relativistic charge on a circular orbit takes the form
\begin{equation}
W_4 = \frac{1}{3} \frac{\omega_0^2 g^2}{\gamma^2} \frac{v_0^2}{(1-v_0^2)^3}.
\end{equation}
Hence, we find the normalised spectral distribution of radiation power of the charge on a circular orbit in four spacetime dimensions as
\begin{equation}
\overline{W}_4^l = 6 \frac{l^{2(l+1)}}{(2l+1)!} (1-v_0^2)^3 v_0^{2(l-1)} {_1}F_2 \left( l + \frac{1}{2}; l + \frac{3}{2}, 1 + 2l; - l^2 v_0^2 \right).
\end{equation}

\begin{figure}[t]
\begin{minipage}[h]{0.49\linewidth}
\center{\includegraphics[width=1\linewidth]{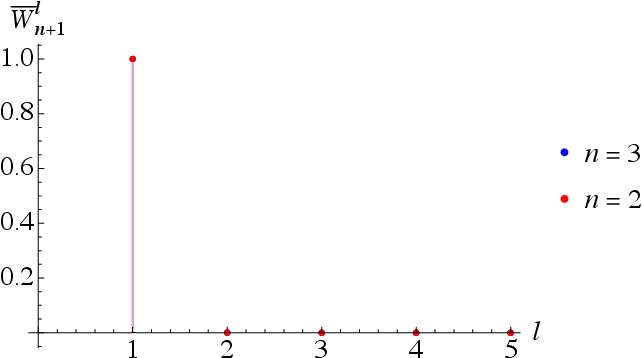}}
\end{minipage}
\hfill
\begin{minipage}[h]{0.49\linewidth}
\center{\includegraphics[width=1\linewidth]{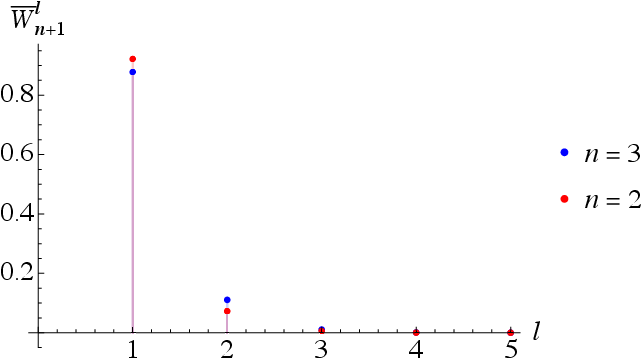}}
\end{minipage}
\begin{minipage}[h]{0.49\linewidth}
\center{\includegraphics[width=1\linewidth]{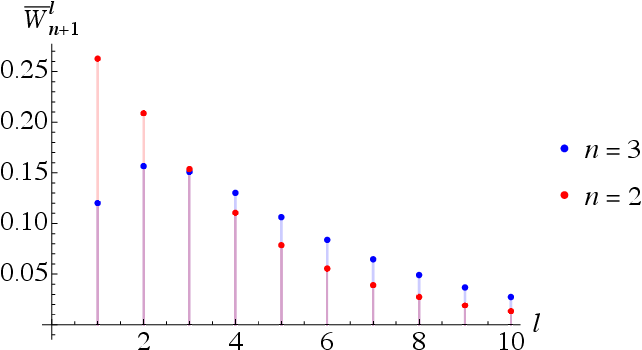}}
\end{minipage}
\hfill
\begin{minipage}[h]{0.49\linewidth}
\center{\includegraphics[width=1\linewidth]{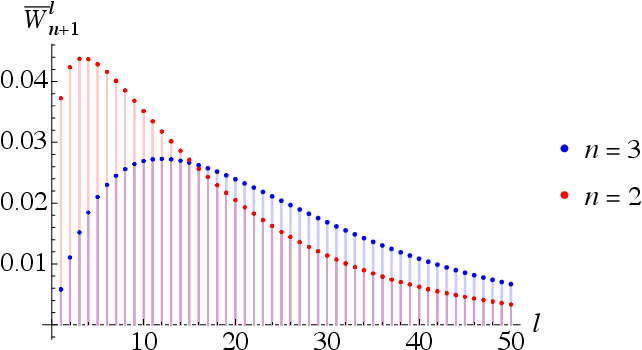}}
\end{minipage}
\caption{Spectral distributions of radiation power of the relativistic charge on a circular orbit in three and four spacetime dimensions for $v_0 = 0.01, \, 0.2, \, 0.7, \, 0.9$ (from the top left to the bottom right).}
\label{fig:6}
\end{figure}

From Fig. \eqref{fig:6} one finds that in three and four dimensions the spectral distributions of radiation power have a similar characteristic form. In the non-relativistic limit, in both cases the main contribution into the radiation power is given by the lowest harmonic of the spectrum $l=1$. The differences between dimensions manifest with the increasing charge velocity. Namely, in three dimensions the maximum of the spectral distribution corresponds to the lower harmonics of the spectrum, compared with its four-dimensional analog. Based on the general formula for the spectral distribution of radiation power in $D=n+1$ dimensions \eqref{eq:n+1_sp_ang_distr}, we conjecture that in higher dimensions the behaviour of spectral distribution should be opposite -- its maximum should correspond to the higher harmonics of the spectrum, compared to the four-dimensional case.

\subsection{Non-relativistic charge on an elliptical orbit}

Let us now consider the second limiting case -- radiation of the non-relativistic charge on elliptical orbit. In the non-relativistic limit
\begin{equation}
|\mathbf{v}| = a \omega_0 \sqrt{\frac{1 + e \cos \xi}{1 - e \cos \xi}} \ll 1 \quad \Longrightarrow \quad v_c \equiv a \omega_0 \ll 1,
\end{equation}
we expand the integral $I_4^l$ up to the contributions linear in $v_c$ as
\begin{align}
I_4^l & = \frac{1}{\omega_0} e^{il v_c e \cos \phi \sin \theta} \int_{0}^{2\pi} d\xi \, (1 - e \cos \xi) e^{il(\xi - e \sin \xi)} \left \lbrack 1 - il v_c \sqrt{1-e^2} \sin \xi \sin \phi \sin \theta \right. \nonumber \\ & - il v_c \cos \xi \cos \phi \sin \theta + {\cal O}(v_c^2) \Big \rbrack.
\end{align}
In what follows, we omit the complex exponent in front of the integral, since we are interested only in the modulus squared of $I_4^l$.

One can easily show that the leading contribution into the integral $I_4^l$ vanishes
\begin{equation}
\int_{0}^{2\pi} d\xi \, (1 - e \cos \xi) e^{il(\xi - e \sin \xi)} = 0.
\end{equation}
Thus, in the non-relativistic limit $I_4^l$ is rewritten as a combination of two integrals
\begin{align}
\label{eq:I_4^l_ellipt_nonrel}
I_4^l & = - i \frac{lv_c}{\omega_0} \sin \theta \left \lbrack \sqrt{1-e^2} \sin \phi \int_{0}^{2\pi} d\xi \, (1 - e \cos \xi) \sin \xi \, e^{il(\xi - e \sin \xi)} \right. \nonumber \\ & + \left. \cos \phi \int_{0}^{2\pi} d\xi \, (1 - e \cos \xi) \cos \xi \, e^{il(\xi - e \sin \xi)} \right \rbrack.
\end{align}
The obtained integrals after some transformations are reduced to the combinations of integrals of the form \cite{Zwillinger_table}
\begin{align}
\label{eq:Bessel_int_1}
&\int_{0}^{\pi} dx \, \sin(nx) \sin(z \sin x) = \frac{\pi}{2} (1 - (-1)^n) J_n(z), \quad n \in \mathbb{Z}, \\
\label{eq:Bessel_int_2}
& \int_{0}^{\pi} dx \, \cos(nx) \cos(z \sin x) = \frac{\pi}{2} (1 + (-1)^n) J_n(z), \quad n \in \mathbb{Z}.
\end{align}
The detailed computation of the integral \eqref{eq:I_4^l_ellipt_nonrel} is relegated to the Appendix \ref{A}. As a result, for two integrals in the Eq. \eqref{eq:I_4^l_ellipt_nonrel} we find
\begin{align}
& \int_{0}^{2\pi} d\xi \, (1 - e \cos \xi) \sin \xi \, e^{il(\xi - e \sin \xi)} = i \pi A_l(e), \\
& \int_{0}^{2\pi} d\xi \, (1 - e \cos \xi) \cos \xi \, e^{il(\xi - e \sin \xi)} = \pi B_l(e),
\end{align}
where we introduced the following notations
\begin{align}
\label{eq:aux_notation_A_l(e)}
& A_l(e) = \left \lbrack J_{l-1}(el) - J_{l+1}(el) - \frac{e}{2} J_{l-2}(el) + \frac{e}{2} J_{l+2}(el) \right \rbrack, \\
& B_l(e) = \left \lbrack J_{l-1}(el) + J_{l+1}(el) - e J_l(el) - \frac{e}{2} J_{l-2}(el) - \frac{e}{2} J_{l+2}(el) \right \rbrack.
\end{align}
Therefore, in the non-relativistic limit the integral $I_4^l$ takes the form
\begin{equation}
I_4^l = \frac{\pi l v_c}{\omega_0} \sin \theta \left \lbrack A_l(e) \sqrt{1-e^2} \sin \phi - i B_l(e) \cos \phi \right \rbrack.
\end{equation}

By analogy with the charge on a circular orbit, we construct the normalised spectral distributions of radiation power of the non-relativistic charge on an elliptical orbit in three and four spacetime dimensions, in accordance with the Eq. \eqref{eq:3D_circ_spect_distr}, as
\begin{align}
\label{eq:nr_ell_orbs}
& \overline{W}_3^l = \frac{l^3 \left \lbrack A_l^2(e) (1-e^2) + B_l^2(e) \right \rbrack}{\sum_{n=1}^{\infty} n^3 \left \lbrack A_n^2(e) (1-e^2) + B_n^2(e) \right \rbrack}, \quad \overline{W}_4^l = \frac{l^4 \left \lbrack A_l^2(e) (1-e^2) + B_l^2(e) \right \rbrack}{\sum_{n=1}^{\infty} n^4 \left \lbrack A_n^2(e) (1-e^2) + B_n^2(e) \right \rbrack}.
\end{align}
Here we estimate the obtained normalising factors numerically.

\begin{figure}[t]
\begin{minipage}[h]{0.49\linewidth}
\center{\includegraphics[width=1\linewidth]{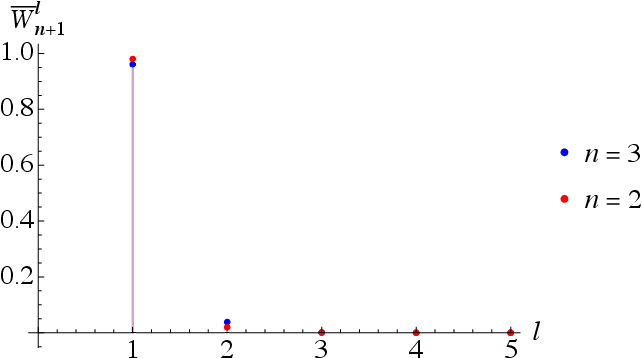}}
\end{minipage}
\hfill
\begin{minipage}[h]{0.49\linewidth}
\center{\includegraphics[width=1\linewidth]{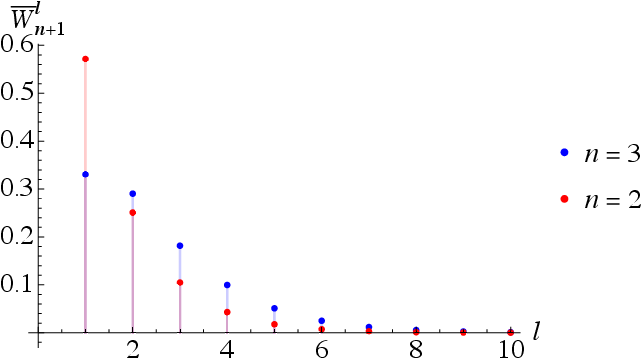}}
\end{minipage}
\begin{minipage}[h]{0.49\linewidth}
\center{\includegraphics[width=1\linewidth]{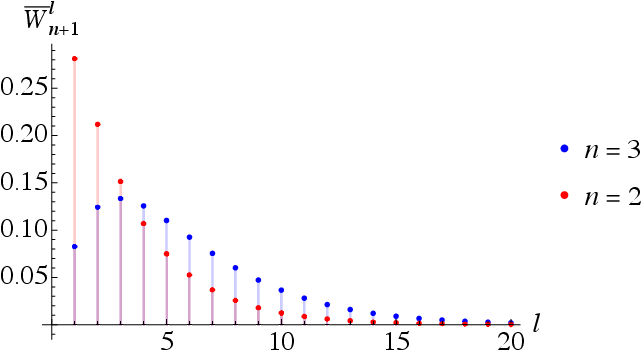}}
\end{minipage}
\hfill
\begin{minipage}[h]{0.49\linewidth}
\center{\includegraphics[width=1\linewidth]{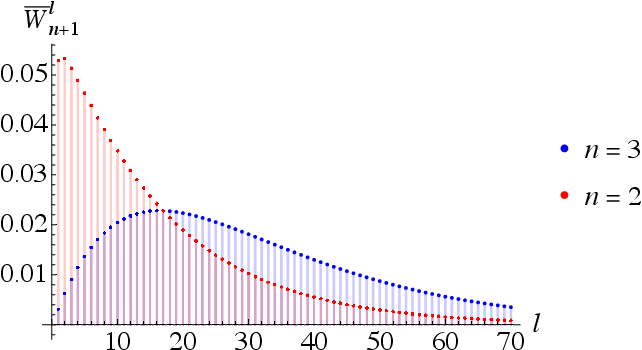}}
\end{minipage}
\caption{Spectral distributions of radiation power of the non-relativistic charge on an elliptical orbit in three and four spacetime dimensions for values of the orbital eccentricity $e= 0.1, \, 0.5, \, 0.7, \, 0.9$ (from the top left to the bottom right).}
\label{fig:7}
\end{figure}

By analogy with the case of the charge on a circular orbit, from Fig. \eqref{fig:7} one finds that in three and four dimensions the spectral distributions of radiation power have a similar characteristic form. In both cases, for orbits with small eccentricity $e \to 0$ the main contribution into the radiation power is given by the lowest harmonic of the spectrum $l=1$. The differences between dimensions manifest themselves with increasing orbital eccentricity. Namely, in three dimensions for orbits with high eccentricity $e \to 1$ the maximum of spectral distribution corresponds to the lower harmonics of the spectrum, compared with the case $D=4$, by analogy with the radiation of ultra-relativistic charge on a circular orbit (see Fig. \eqref{fig:6}). Also, based on the Eqs. \eqref{eq:n+1_arbit_spectrum} and \eqref{eq:nr_ell_orbs} we conjecture that in higher dimensions the behaviour of spectral distributions is opposite -- the maximum of spectral distribution corresponds to the higher harmonics, compared to the four-dimensional case.

\section{Conclusions}\label{V}

In this work, our goal was to study the non-local effects in gravitational radiation of binary systems in theories with odd number of non-compact extra spacetime dimensions, associated with the Huygens principle violation. We considered a simple model of the scalar field interacting with the point charge moving along a fixed elliptical orbit in three-dimensional Minkowski spacetime. Although this model is physically non-viable, it provides us with a simple framework to study the effects resulting from the Huygens principle violation, which are expected in the gravitational radiation of elliptical binary systems in realistic gravity models with odd number of non-compact extra dimensions such as the RS2 and DGP models.

We demonstrated that the Huygens principle violation in three dimensions results into the formation of a characteristic tail signal in the radiation of the charge. In particular, it manifests in the shifts of extremum points of the charge radiation power in time from the moments when the charge passes the pericenter and apocenter of the orbit. Also, the spectral distribution of the radiation power of the charge is obtained. It is found that in three spacetime dimensions the spectral distribution of radiation power of the charge on elliptical orbit has a characteristic form similar to its four-dimensional analog. However, its maximum corresponds to the lower harmonics of the spectrum, compared to the four-dimensional case.

We consider the scalar field radiation by the particle moving along a fixed elliptical orbit in three spacetime dimensions neglecting the radiation reaction on its orbit. We extract the emitted part of the three-dimensional scalar field \eqref{eq:2+1_emit_part} by use of the Rohrlich-Teitelboim approach to radiation and obtain the formula for the radiation power of a non-relativistic charge moving along an arbitrary trajectory \eqref{eq:2+1_nr_rad_pow}. In accordance with the Huygens principle violation in odd dimensions, it depends on the entire history of charge motion preceding the retarded time. Using a simple example of the non-relativistic charge with acceleration depending on time as the Gaussian function \eqref{eq:Gauss_acc}, we demonstrate the presence of tail signals in radiation of non-relativistic localised sources in odd spacetime dimensions (see Fig. \eqref{fig:1}). For the charge on elliptical orbit, we calculate the integral over the history of its motion up to the contributions quadratic in the orbital eccentricity \eqref{eq:hist_int_ellipt_2nd} and find the corresponding dependence of the charge radiation power on time \eqref{eq:Rad_pow_2nd}. In this case, the tail term manifests in radiation due to the shifts of the radiation power extremum points in time from the moments when the charge passes the pericenter and apocenter of the orbit (see Figs. \eqref{fig:4} and \eqref{fig:5}). We obtain the formulae for these shifts up to the contributions quadratic in the orbital eccentricity \eqref{eq:extr_pos}.

We also study the spectral distribution of radiation power of the charge on elliptical orbit in three dimensions. We consider two limiting cases -- the relativistic charge on a circular orbit \eqref{eq:2+1_circ_spectrum} and the non-relativistic charge on an elliptical orbit \eqref{eq:nr_ell_orbs}. In both cases, we find the characteristic feature of three-dimensional radiation: in three dimensions the maximum of the spectral distribution corresponds to the lower harmonics of the spectrum, compared to the four-dimensional case (see Fig. \eqref{fig:6} and \eqref{fig:7}). Based on the Eqs. \eqref{eq:n+1_arbit_spectrum}, \eqref{eq:2+1_circ_spectrum} and \eqref{eq:nr_ell_orbs} we conjecture that in higher dimensions behaviour of the spectral distributions is opposite -- their maximums should correspond to the higher harmonics of the spectrum, compared to the four-dimensional theory.

The analogous tail signals in radiation in odd dimensions were found in Refs. \cite{Barvinsky:2003jf} and \cite{Chu:2021uea} for simple examples of radiation sources. Namely, within the framework of a scalar field analog of the RS1 model, in Ref. \cite{Barvinsky:2003jf} the presence of the tail signal in radiation of a scalar charge living a finite time interval was demonstrated. Also, in Ref. \cite{Chu:2021uea} the tail signal was found in the odd-dimensional electromagnetic and gravitational radiation of a harmonically oscillating source. In our work, we have demonstrated the presence of the tail signal in radiation of a realistic astrophysical source of gravitational waves in odd spacetime dimensions.

\section*{Acknowledgements}

This work is supported by the Foundation for the Advancement of Theoretical Physics and Mathematics “BASIS” grant No. 20-2-10-8-1. The author would like to thank D.V.~Gal'tsov for valuable discussions and proofreading of the paper.

\appendix
\section{Calculation of $I_4^l$ for a charge on elliptical orbit}\label{A}

The spectral distribution of radiation power of the charge on elliptical orbit is determined by the integral \eqref{eq:I_4^l_ellipt_nonrel}, which is given by a linear combination of two integrals
\begin{align}
\label{eq:aux_int_sin}
& I_{\rm 4,s}^{l} = \int_{0}^{2\pi} d\xi \, (1 - e \cos \xi) \sin \xi \, e^{il(\xi - e \sin \xi)}, \\
\label{eq:aux_int_cos}
& I_{\rm 4,c}^{l} = \int_{0}^{2\pi} d\xi \, (1 - e \cos \xi) \cos \xi \, e^{il(\xi - e \sin \xi)}.
\end{align}
Here we demonstrate only the calculation of the integral \eqref{eq:aux_int_sin}. The integral \eqref{eq:aux_int_cos} is computed in a completely analogous way.

Given that the integral of an odd periodic function over the period vanishes, expanding the exponent by use of the Euler formula we arrive at the following integral
\begin{equation}
I_{\rm 4,s}^{l} = i \int_{0}^{2\pi} d\xi \, (1 - e \cos \xi) \sin \xi \sin(l\xi - el \sin \xi).
\end{equation}
Introducing new integration variable $\xi' = \xi - \pi$, we rewrite $I_{\rm 4,s}^{l}$ as
\begin{equation}
I_{\rm 4,s}^{l} = i (-1)^{l+1} \int_{-\pi}^{\pi} d\xi' \, (1 + e \cos \xi') \sin \xi' \sin(l\xi' + el \sin \xi').
\end{equation}
Since we have obtained the integral of an even function over a symmetric interval, expanding $\sin(l\xi' + el \sin \xi')$ and renaming the integration variable $\xi' \to \xi$ we find $I_{\rm 4,s}^{l}$ in the form
\begin{align}
I_{\rm 4,s}^{l} & = 2i (-1)^{l+1} \int_{0}^{\pi} d\xi \, \Big \lbrack \sin \xi \sin{l\xi} \cos(el \sin \xi) + \sin \xi \cos{l\xi} \sin(el \sin \xi) \nonumber \\ & + \frac{e}{2} \sin{2\xi} \sin{l\xi} \cos(el \sin \xi) + \frac{e}{2} \sin{2\xi} \cos{l\xi} \sin(el \sin \xi) \Big \rbrack.
\end{align}
Finally, representing the products of type $\sin{m\xi} \cos{l\xi}$ as the combinations of sines/cosines of the sum or difference we obtain
\begin{align}
I_{\rm 4,s}^{l} & = i (-1)^{l+1} \int_{0}^{\pi} d\xi \, \Big \lbrack \cos{(l-1)\xi} \cos(el \sin \xi) - \cos{(l+1)\xi} \cos(el \sin \xi) \nonumber \\ & - \sin{(l-1)\xi} \sin(el \sin \xi) + \sin{(l+1)\xi} \sin(el \sin \xi) + \frac{e}{2} \cos{(l-2)\xi} \cos(el \sin \xi) \nonumber \\ & - \frac{e}{2} \cos{(l+2)\xi} \cos(el \sin \xi) - \frac{e}{2} \sin{(l-2)\xi} \sin(el \sin \xi) + \frac{e}{2} \sin{(l+2)\xi} \sin(el \sin \xi) \Big \rbrack.
\end{align}
The resulting integrals are calculated by the Eqs. \eqref{eq:Bessel_int_1} and \eqref{eq:Bessel_int_2} and we find the following expression for the integral \eqref{eq:aux_int_sin}
\begin{equation}
I_{\rm 4,s}^{l} = i \pi A_l (e),
\end{equation}
where $A_l(e)$ is given by the Eq. \eqref{eq:aux_notation_A_l(e)}.

\end{document}